\documentclass[reprint,english,aps,prb,raggedbottom,showpacs,floatfix]{revtex4-1}

\usepackage[T1]{fontenc}
\usepackage[utf8]{inputenc}
\usepackage{babel}
\usepackage{verbatim}
\usepackage{amsmath}
\usepackage{amssymb}
\usepackage{graphicx}
\usepackage{esint}
\usepackage[colorlinks=true
  ,linkcolor=blue
  ,citecolor=blue
  ,urlcolor=blue
  ,allcolors=blue]{hyperref}

\usepackage{breakurl}
\usepackage[normalem]{ulem}

\usepackage{color}
\usepackage{soul}

\usepackage[table]{xcolor}

\usepackage{stackrel}

\begin{document}

\title{Orbital-selective bad metals due to Hund's rule and orbital anisotropy: a finite-temperature slave-spin treatment of the two-band Hubbard model}

\author{Alejandro Mezio}
\email{alejandro.mezio@gmail.com}
\affiliation{School of Mathematics and Physics, University of Queensland, Brisbane QLD 4072, Australia}

\author{Ross H. McKenzie}
\email{r.mckenzie@uq.edu.au}
\affiliation{School of Mathematics and Physics, University of Queensland, Brisbane QLD 4072, Australia}

\begin{abstract}
We study the finite-temperature properties of the half-filled two-band Hubbard model in the presence of Hund's rule coupling and orbital anisotropy.
We use the mean-field treatment of the $Z_2$ slave-spin theory with a finite-temperature extension of the zero-temperature gauge variable previously developed by Hassan and de' Medici [\href{http://dx.doi.org/10.1103/PhysRevB.81.035106}{Phys. Rev. B {\bf 81}, 035106 (2010)}].
We consider the instability of the Fermi liquid phases and how it is enhanced by the Hund's rule.
We identify paramagnetic solutions that have zero quasi-particle weight with bad metallic phases, and the first-order transition temperature between it and the Fermi liquid phase as a coherence temperature that signals the crossover to the bad metallic state.
When orbital anisotropy is present, we found an intermediate transition to an orbital-selective bad metal (OSBM), where the narrow band becomes a bad metal while the wide band remains a renormalised Fermi liquid.
The temperatures $T_\textrm{coh}$ and $T_\textrm{OSBM}$ at which the system transitions to the bad metal phases can be orders of magnitude less than the Fermi temperature associated with the non-interacting band.
The parameter dependence of the temperature at which the OSBM is destroyed can be understood in terms of a ferromagnetic Kondo-Hubbard lattice model.
In general, Hund's rule coupling enhances the bad metallic phases, reduce interorbital charge fluctuations and increase spin fluctuations.
The qualitative difference found in the ground state whether the Hund's rule is present or not, related to the degeneracy of the low energy manifold, is also maintained for finite temperatures.
\end{abstract}


\maketitle

\section{Introduction}

One of the most interesting new ideas about quantum matter from the last decade is that of a Hund's metal.\cite{Yin2011,Haule2009a,Georges2013a,DeMedici2017b}
This is a strongly correlated metal that can occur in a multi-orbital material as a result of the Hund's rule interaction $J$, that favours parallel spins in different orbitals.
While strong correlation effects are often associated with the proximity to a Mott insulating state, it has become clear in recent years that the Hund's rule coupling (rather than the Hubbard $U$) is responsible for strong correlations in multiorbital metallic materials that are not close to a Mott insulator, such as the iron-based superconductors,\cite{Haule2009a,Yin2011} and ruthenates.\cite{Werner2008,Mravlje2011}
Besides the enhanced electron correlation, this new type of strongly correlated system is characterised by local high spin configurations with slow dynamics and selectivity of the electron correlations depending on the orbital character.\cite{Werner2008,Haule2009a,DeMedici2017b}
Hund's coupling considerably reduces the low-energy quasiparticle coherence scale, that results in an incoherent metallic state with frozen local moments in an extended temperature range above it, i.e., a bad metal.

It has been shown that Hund's rule has a conflicting effect on the correlations of multiorbital systems.
At integer fillings, its modifies the critical interaction where the metal-insulator transition (MIT) occurs, $U_\textrm{MIT}$, depending on the number of electrons per site,\cite{DeMedici2011a,Georges2013a} and reduces the temperature scale $T_\textrm{coh}$ above which a bad metal is formed.\cite{Haule2009a,Ong2012}
In the one band Hubbard model, this coherence temperature $T_\textrm{coh}$ is orders of magnitude smaller than the bare energy scales of the system ($U$ and bandwidth $W$) and signals the breakdown of the low-temperature Fermi liquid (FL) picture and the crossover to a bad metal state.
Several other signatures of this FL to bad metal crossover at $T_\textrm{coh}$ exist:
the resistivity becomes of order of the Mott-Ioffe-Regel limit ($\frac{h\, a}{e^2}\sim 0.1\, m\Omega cm$), an incoherent electron spectral function, a collapse of the Drude peak in the optical conductivity and a shift of the associated spectral weight to higher energies, the entropy and specific heat become of order $k_B$ per particle, the NMR Knight shift dependence with the temperature becomes consistent with a local-moments dominated behaviour (Curie-Weiss), and sometimes there is a nonmonotonic temperature dependence of the Hall coefficient and thermoelectric power.\cite{Rozenberg1995,Merino2000,Gunnarsson2003,Hussey2004a,Merino2008,Deng2013a,Xu2013,Vucicevic2013,Vucicevic2015,Dasari2016b}
Usually associated with the proximity to a Mott MIT, it is interesting to ask how and why the Hund's rule interaction and orbital degeneracy and character change this low-temperature crossover and enhance the formation of bad metals.

When orbitals have different bandwidths or their degeneracy is lifted by a crystal field, correlations can affect each band differently.
Some orbital-dependent correlations have been investigated in theoretical calculations for iron-based superconductors\cite{DeMedici2009,Ishida2010,Yu2011,Bascones2012,Yu2012,Georges2013a,Yi2013,Yu2013,Terashima2013,DeMedici2014,Liu2015,Yi2015} and ruthenates.\cite{Anisimov2002,Koga2004,DeMedici2005,Mravlje2011,Georges2013a}
Hund's rule decouples the orbitals, enhancing such orbital differentiation,\cite{DeMedici2005,DeMedici2009,DeMedici2011,Yu2013} and an extreme case occurs at $T=0$ when some orbitals transition to a Mott phase while others remain metallic, leading to an orbital-selective Mott phase (OSMP).\cite{Anisimov2002,Liebsch2004,Koga2004,DeMedici2005,Ferrero2005,Inaba2007a}
The two-band Hubbard-Kanamori model with unequal bandwidths is the simplest model where a transition to an OSMP occurs,\cite{Anisimov2002,Koga2004,DeMedici2005,Ferrero2005,Liebsch2005,Biermann2005,Werner2007a,Vojta2010} and some earlier numerical works using dynamical mean-field theory (DMFT) explore its effects at finite temperature.\cite{Liebsch2004,Biermann2005,Knecht2005,Liebsch2005,Inaba2005,Liebsch2006,Inaba2007}

Based on scanning tunnelling microscope (STM), recent quasi-particle interference measurements of the normal state Fermi surface and superconducting energy gaps on FeSe, give support to the idea that orbital-selective strong correlations dominate the parent state of iron-based superconductors.\cite{Sprau2017a,Kostin2018}
Including these values of orbital-selective quasiparticle weights into a spin-fluctuation pairing theory in a random-phase approximation (RPA) study, some of the authors of the previous papers obtain an accurate description for the superconducting gap, indicating the key role of orbital-selective Cooper pairing.\cite{Sprau2017a,Kreisel2017}
And more recently a good agreement in the calculated magnetic excitation spectrum with inelastic neutron scattering experiments in FeSe.\cite{Kreisel2018}

Angle-resolved photoemission spectroscopy studies on several iron-based superconductors show a temperature-induced crossover from a metallic FL state at low temperature with well-defined Fermi surfaces for all the bands, to a phase where the $d_{xy}$ orbital loses spectral weight with increasing temperature and the associated Fermi surface dissapears.
See Ref. \onlinecite{Yi2013} for results in $A_x$Fe$_{2-y}$Se$_2$ ($A$ = K, Rb);
Ref. \onlinecite{Yi2015} for FeTe$_{0.56}$Se$_{0.44}$, K$_{0.76}$Fe$_{1.72}$Se$_2$ and FeSe grown on SrTiO$_3$;
Ref. \onlinecite{Liu2015} for Fe$_{1+y}$Se$_x$Te$_{1-x}$ ($0<x<0.59$);
Ref. \onlinecite{Pu2016} for single layer FeSe/Nb:BaTiO$_3$/KTaO$_3$;
and Ref. \onlinecite{Miao2016a} for LiFeAs.
These results indicate an orbital-differentiated coherent-incoherent crossover, and are consistent with a scenario of a Kondo-type screening determined by the strength of the Hund's rule coupling.
Other experimental probes also find a coherent-incoherent crossover, with signatures of a bad metal behaviour: temperature dependence of the Knight shift consistent with a Curie-Weiss behaviour\cite{Wu2016} and strong temperature dependence of the Hall coefficient\cite{Xiang2016a} in $A$Fe$_2$Se$_2$ ($A$ = K, Rb, Cs), and a collapse of the Drude peak in the optical conductivity of KFe$_2$Se$_2$, where spectral weight is transferred from low to high energy.\cite{Yang2017a}

An important question concerns the extent to which slave-particle mean-field theories can capture the stability of the Hund's metal and its properties, including the emergence of a bad metal above some coherence temperature, $T_\textrm{coh}$.
In the single-band Hubbard model, the strongly correlated metallic phase that occurs in proximity to a Mott MIT is associated with a small quasi-particle weight and suppression of double occupancy, reflecting suppressed charge fluctuations.
This is captured by slave-boson mean-field theory, including the small coherence temperature.\cite{Dao2017,Mezio2017}
In contrast, the strongly correlated Hund's metal is associated with suppression of singlet spin fluctuations on different orbitals, without suppression of onsite charge fluctuations, and is seen with the $Z_2$ slave-spin mean-field (SSMF) theory at zero temperature.\cite{Fanfarillo2015,DeMedici2017,DeMedici2017a}
To the best of our knowledge, there is no work studying an extension of this $Z_2$ SSMF theory to finite temperatures, but only using other variants of the method.\cite{Yu2013d,Yi2013,Gao2014,Yi2015,Yang2018}

\begin{figure}
  \includegraphics*[width=0.46\textwidth]{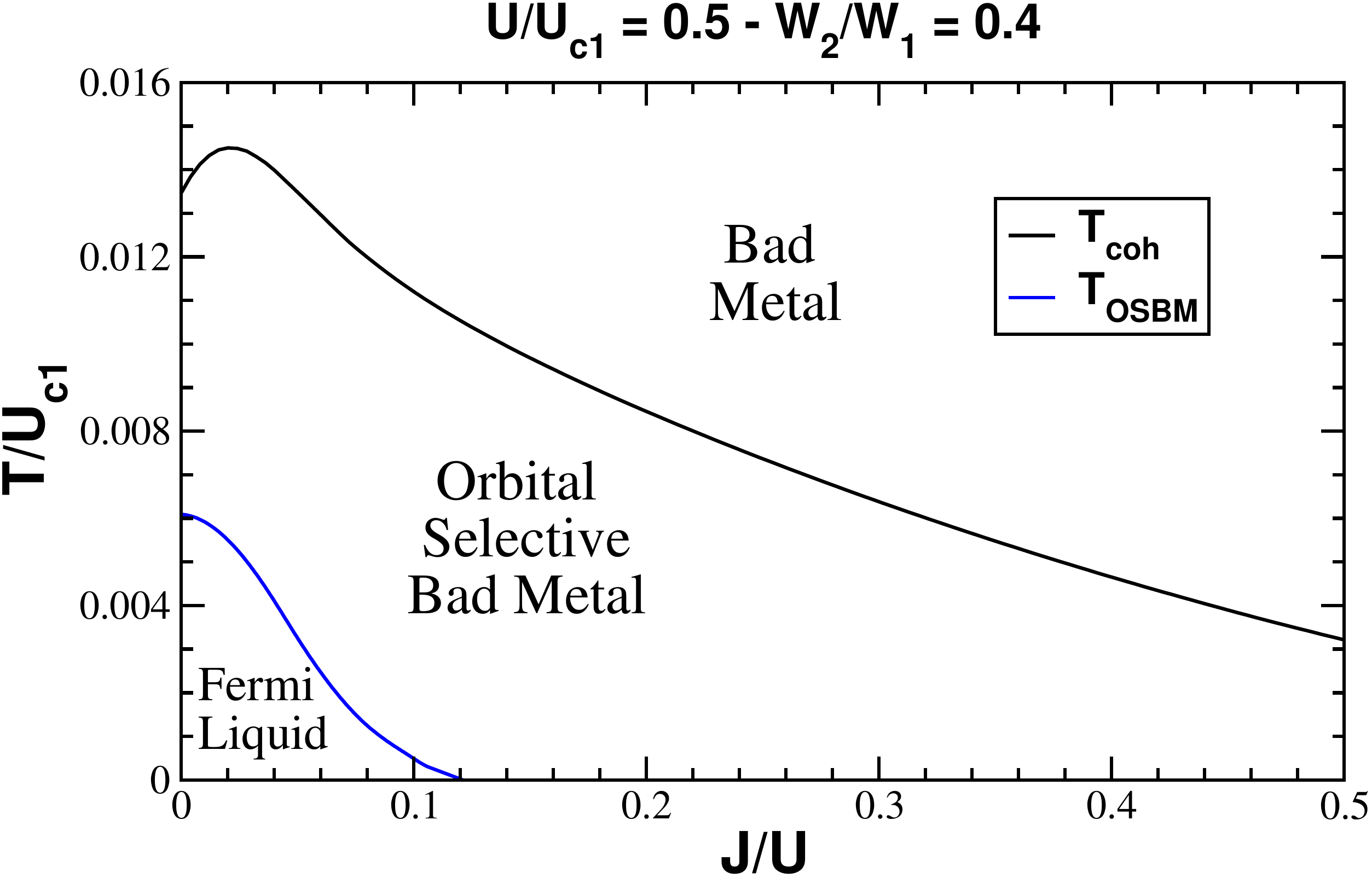}
  \caption{\label{fig:main-result} Phase diagram for $T$ vs $J/U$.
  Stabilisation of the orbital-selective bad metal by Hund's rule interaction.
  The system is at half-filling, the interaction strength is $U/U_\textrm{c1}=0.5$ and orbital anisotropy $W_2/W_1=0.4$.}
\end{figure}

In this paper, we propose a finite-temperature implementation of the $Z_2$ SSMF theory that is a natural extension of the $T=0$ formulation,\cite{Hassan2010,DeMedici2017a} and we apply it to the two-band Hubbard-Kanamori model at half-filling.
We explore the effects of the Hund's rule $J$ and orbital anisotropy in the coherence temperature $T_\textrm{coh}$, and the inter-orbital spin and charge fluctuations.
We also investigate the appearance of an \emph{orbital-selective bad metal} phase, where one band has incoherent quasiparticles, i.e., bad metal, while the other remains a FL.
Fig. \ref{fig:main-result}
shows the phase diagram for $T$ vs $J/U$ for the two-band Hubbard Kanamori model with different orbital bandwidths and at intermediate interaction $U$, and summarises our main result.
The Hund's rule interaction enhances the stability of the orbital-selective bad metal phase, strongly reducing the (first) coherence temperature $T_\textrm{OSBM}$. It also increases correlations, reducing the (second) coherence temperature $T_\textrm{coh}$, where the remaining metallic band becomes a bad metal.
At $T=0$, a transition from the FL to an OSMP occurs at a critical value of the Hund's coupling $J/U \simeq 0.12$.
In this way, even for low-$J$ the system is close to an orbital-selective Mott phase, and an increase in temperature favours the OSBM phase. This occurs even in the case of small anisotropy (see Fig. \ref{fig:aniso-Tcoh-moveW}).

It is worth clarifying that previous SSMF studies at $T=0$ identify the slave-spin paramagnetic phase where the quasiparticle weight is $Z=0$ with a Mott insulator phase, while it has been pointed out that beyond the single-site approximation this might be an orthogonal metal, a type of fractionalised non-Fermi liquid.\cite{Nandkishore2012}
In our paper we use the single-site mean-field approximation (valid in the large dimension limit). Although our results in the slave-spin  paramagnetic phase can suggest a complex behaviour (e.g., see Sec. \ref{sec:entropy}), the temperature dependence of this phase follows a simple thermal activation of the atomic slave-spin states.

The organization of the paper is as follows: In Sec. \ref{sec:model-method}, we describe the Hubbard-Kanamori model and SSMF method.
Details on the finite-$T$ implementation are in Appendix \ref{opc-append}.
In Sec. \ref{sec:results}, we present our results for the temperature dependence of the quasiparticle weight, coherence temperature, spin and charge fluctuations, phase diagrams and entropy contributions.
We identify the first-order transition temperature where the quasiparticle weight $Z$ vanishes to the coherence temperature $T_\textrm{coh}$ associated with the crossover to a bad metal.
The behaviour of $T_\textrm{coh}$ is qualitatively different whether the Hund's rule coupling is present or not, as it is found for the ground state and explained from the degeneracy of the low energy manifold.\cite{Koga2004,DeMedici2017a}
We found that the change of $T_\textrm{coh}$ when moving $J/U$ occurs only through the modification of the zero-$T$ critical interaction where the MIT occurs, $U_\textrm{MIT}$.
When orbital anisotropy is present the width of the two bands are unequals (i.e. $W_1\neq W_2$), and we found that the Hund's rule facilitates the first-order transition from the FL to a state where the narrow band quasiparticle weight vanishes.
We identify this with a crossover to an orbital-selective bad metal state and this additional coherence temperature as $T_\textrm{OSBM}$.
Details on the construction of the solutions and an analysis of the atomic states are in Appendices \ref{solutions-append} and \ref{atomic-append}, respectively.

\section{Model and Method}
\label{sec:model-method}

\subsection{Model Hamiltonian}
Our starting point is the general multi-band Hubbard-Kanamori Hamiltonian\cite{Georges2013a} which describes interacting electrons in $N_o$ orbitals,
\begin{equation}
\hat{\mathcal{H}}= \hat{\mathcal{H}}_0 + \sum_i \left( \hat{\mathcal{H}}_{U} + \hat{\mathcal{H}}_{J} \right)  - \mu\, \hat{N}\ ,
\label{Hamiltonian1}
\end{equation}
where,
\begin{equation}
\hat{\mathcal{H}}_0 =  \sum_{\substack{i \neq j,\sigma \\ m,m'}} t^{mm'}_{ij} \hat{c}^\dag_{im\sigma} \hat{c}_{jm' \sigma} + \sum_{i,m,\sigma} \epsilon_m \hat{n}_{im\sigma}\ ,
\end{equation}
is the non-interacting term, and $\hat{N}=\sum_i \hat{n}_i$ the total number of electrons.
As usual, $\hat{c}^\dag_{im\sigma}$ creates an electron with spin $\sigma=\, \uparrow, \downarrow$ at the site $i=1,\dots,N_s$ on the orbital $m=1, \dots, N_o$, and $\hat{n}_i=\sum_{m,\sigma} \hat{c}^\dag_{im\sigma} \hat{c}_{im\sigma}$ is the occupation for the site $i$.
The hopping matrix element $t^{mm'}_{ij}$ satisfies $t^{mm'}_{ij}=\left(t^{m' m}_{ji}\right)^*$, has no inter-orbital hybridisation ($t^{mm'}_{ij}=\delta_{mm'} t^{m}_{ij}$), and we write out explicitly the orbital energies  ($t^{m}_{ii}=\epsilon_m$).
The interaction terms are,
\begin{align}
\hat{\mathcal{H}}_{U} =&\ U \sum_m \hat{n}_{m\uparrow} \hat{n}_{m\downarrow} + U' \sum_{m\neq m'} \hat{n}_{m\uparrow} \hat{n}_{m'\downarrow} \nonumber \\
&\ + \left( U' - J \right) \sum_{m< m',\sigma} \hat{n}_{m\sigma} \hat{n}_{m'\sigma}\ , \label{HU}\\
\hat{\mathcal{H}}_{J} =&\ - J_X \sum_{m\neq m'} \hat{c}^\dag_{m\uparrow} \hat{c}_{m\downarrow} \hat{c}^\dag_{m' \downarrow} \hat{c}_{m' \uparrow} \nonumber \\
&\ + J_P \sum_{m\neq m'} \hat{c}^\dag_{m\uparrow} \hat{c}^{\dag}_{m\downarrow} \hat{c}_{m' \downarrow} \hat{c}_{m' \uparrow} \label{HJ}
\end{align}
where we omit the site label $i$.
The density-density term $\hat{\mathcal{H}}_{U}$ involves the on-site Coulomb interaction between electrons in the same orbital with opposite spins $U$, in different orbitals with opposite spins $U'$, and different orbitals with parallel spins $U'-J$; while $\hat{\mathcal{H}}_{J}$ involves the \emph{spin-flip} ($J_X$) and \emph{pair-hopping} ($J_P$) interactions.

The model has rotational symmetry whether one chooses $J_X=J_P=J$ and $U'=U-2J$, or sets $J_P=0$, $J_X=J$ and $U'=U-J$.
Although the former case refers to the physical Hamiltonian for $t_{2g}$ states, the choice does not affect the results qualitatively and for simplicity we use the latter set of parameters.\cite{Georges2013a,Komijani2017}
We restrict our results to the two band case ($N_o=2$) at half-filling ($n_1=n_2=1$), each with orbital energy $\epsilon_m=0$, bandwidth $W_m$ and a semicircular density of states,
\begin{equation}
\rho_m(\varepsilon) = \frac{8}{\pi}\, \frac{1}{W_m^2}\, \sqrt{\left( \frac{W_m}{2} \right)^2 -\varepsilon^2}\ \ \ \ \ \ \left( m=1,2 \right).
\end{equation}
Several previous works studied the two-band model at zero temperature with the slave-spin method, calculating its different phases and its dependence with Hund's coupling, orbital anisotropy, crystal field splitting, orbital hybridisation, and different fillings.\cite{DeMedici2005,Hassan2010,DeMedici2011,Komijani2017,DeMedici2017a}
We use those results as a guide to benchmark our results when $T=0$, and focus on how $J$, $U$ and the orbital anisotropy $W_2/W_1$ affects the system at finite temperature.

Throughout the paper we use the one-band critical interaction as the energy unit,
\begin{equation}
U_\textrm{c1} \equiv-16\, \overline{\varepsilon}_{(T=0)} = \frac{16}{3\, \pi}\, W_1\ ,
\end{equation}
which is the only relevant energy scale for the one band case at the mean field level.\cite{Mezio2017}
It depends only on the zero-$T$ uncorrelated kinetic energy $\overline{\varepsilon}_{(T=0)}$, that we define later on Eq. \eqref{energy-uncorr}.
An extension of this energy scale for generic filling is $U^*_n = \frac{-16}{n\, (2-n)}\, \overline{\varepsilon}_{(T=0)}$.\cite{Mezio2017}

\subsection{Slave-spin mapping}
\label{sec:slave-spin}

We use the slave-spin mean-field (SSMF) method \cite{DeMedici2005,Hassan2010,Georgescu2015,DeMedici2017a} to study the finite temperature behaviour of the Hubbard-Kanamori Hamiltonian \eqref{Hamiltonian1}.
This method involves a slave-particle representation where we express the physical electron as a product of a fermion and slave spin-$1/2$ operator, allowing a rewriting of the Hamiltonian more suitable for further mean-field approximations. Here the slave-spin states ``up'' or ``down'' labels occupied or unoccupied electronic states, respectively.
Within this mapping, the $\hat{c}_{im\sigma}$ operators in the non-diagonal  part of $\hat{\mathcal{H}}_0$ are replaced by,
\begin{equation}
\hat{c}_{im\sigma}=\hat{f}_{im\sigma}\, \hat{O}_{im\sigma}\label{ss-map} \ ,
\end{equation}
where $\hat{f}_{im\sigma}$ is an auxiliary fermion operator, and $\hat{O}_{im\sigma}$ is a generic slave-spin operator. Its general form is,
\begin{equation}
\hat{O}_{im\sigma} = \left(  \begin{array}{cc}
0 & c_{im\sigma} \\
1 & 0
\end{array} \right)\ ,
\end{equation}
with $c_{im\sigma}$ an arbitrary complex number that we can tune after an approximation scheme to reproduce solvable limits of the problem.
At $T=0$ and single-site mean-field level, a set of choices for this parameter that recover the physical solution in the uncorrelated limit $U=U'=J=0$ and work for generic filling,\cite{Hassan2010,DeMedici2017a} give interesting results and have been thoroughly tested against DMFT, slave-boson and Gutzwiller approximations.\cite{DeMedici2005,DeMedici2009,Hassan2010,DeMedici2011,DeMedici2014,Fanfarillo2015,DeMedici2017a}
However, it fails to satisfy the non-interacting limit when used at finite temperature. In this paper, we develop an extension of this choice of (real) $c$-parameter suitable for finite temperatures.
Details of the calculation are in Appendix \ref{opc-append}.

For the application of the slave-spin mapping on the other terms of the Hamiltonian \eqref{Hamiltonian1}, it is convenient to rewrite the density-density Hamiltonian in a particle-hole symmetric form.
For this, we shift all the number operators in $\hat{\mathcal{H}}_U$ by $1/2$, $\hat{n}_{im\sigma} \rightarrow \hat{n}_{im\sigma}-\frac{1}{2}$.
By doing this, we only add a one-body term that shifts the chemical potential, $\mu \rightarrow \mu -E_0$, and a constant total energy shift, $\hat{\mathcal{H}} \rightarrow \hat{\mathcal{H}} - \frac{N_s\, N_o}{2} E_0$, where,
\begin{equation}
E_0 = \frac{U + U' (N_o -1) + (U'-J) (N_o -1) }{2} = \frac{3\, U}{2}-\frac{3\, J}{2}\label{E0}\ ,
\end{equation}
for two bands and our choice of parameters.
These displaced number operators in $\hat{\mathcal{H}}_U$ have the same quantum numbers and are mapped to the $z$-component slave-spins operators, $\hat{n}_{im\sigma}=\hat{S}^z_{im\sigma}$.
The electron number operators that are not shifted by $1/2$, i.e., those accompanying $\mu$ and $\epsilon_m$, are represented by the auxiliary fermion occupation operator, $\hat{n}_{im\sigma}=\hat{n}^f_{im\sigma}=\hat{f}^\dag_{im\sigma} \hat{f}_{im\sigma}$.
Finally, $\hat{\mathcal{H}}_J$ in Eq. \eqref{HJ} mixes the Hilbert spaces of the $f$ fermions and slave-spins, and we use here the approximate mapping $\hat{c}^\dag_{im\sigma}=\hat{S}^+_{im\sigma}$ and $\hat{c}_{im\sigma}=\hat{S}^-_{im\sigma}$, which has the correct slave-spin quantum numbers and captures the spin-flip and pair-hopping physics in that Hilbert space.\cite{DeMedici2005,DeMedici2017a}

Due to the increase in the size of the Hilbert space by slave-particle methods, constraints must be introduced to reproduce the physical states by the auxiliary ones. In the slave-spin formulation we only need one constraint equation per introduced slave-spin degree of freedom, namely
\begin{equation}
\hat{f}^\dag_{im\sigma}\, \hat{f}_{im\sigma} = \hat{S}^z_{im\sigma} + \frac{1}{2} \label{constraint1}\ .
\end{equation} 

\subsection{Mean-field approximation}
\label{sec:mean-field}

Following Reference [\onlinecite{DeMedici2017a}], we perform a mean-field decoupling for each site between the fermionic and slave-spin degrees of freedom, $\hat{O}^\dag \hat{O} \hat{f}^\dag \hat{f} \simeq \langle \hat{O}^\dag \hat{O} \rangle \hat{f}^\dag \hat{f} + \hat{O}^\dag \hat{O} \langle \hat{f}^\dag \hat{f} \rangle - \langle \hat{O}^\dag \hat{O}\rangle \langle \hat{f}^\dag \hat{f} \rangle$, followed by a single-site mean-field in the slave-spin $\hat{O}$ operators, $\hat{O}^\dag \hat{O} \simeq \langle \hat{O}^\dag \rangle \hat{O} + \hat{O}^\dag \langle \hat{O} \rangle - \langle \hat{O}^\dag \rangle \langle \hat{O} \rangle$.
The constraints in Eqs. \eqref{constraint1} are included through a Lagrange multiplier $\lambda_{im\sigma}$ by adding the term $\sum_{im\sigma} \lambda_{im\sigma} \left( \hat{S}^z_{im\sigma} + \frac{1}{2} - \hat{n}^f_{im\sigma} \right)$ to the Hamiltonian \eqref{Hamiltonian1} .
Finally, we assume translational invariance ($\hat{O}_{im\sigma}=\hat{O}_{m\sigma}$, $\lambda_{im\sigma}=\lambda_{m\sigma}$ and $t^m_{ij}=t^m_{\textbf{R}_j-\textbf{R}_i}$) and paramagnetic solutions ($\langle \hat{O}_{m\uparrow} \rangle=\langle \hat{O}_{m\downarrow} \rangle$, $\lambda_{m\sigma}=\lambda_m$ and $\langle \hat{f}^\dag_{{\bf k}m\uparrow}\, \hat{f}_{{\bf k}m\uparrow} \rangle=\langle \hat{f}^\dag_{{\bf k}m\downarrow}\, \hat{f}_{{\bf k}m\downarrow} \rangle$).

After the mean-field approximations and assumptions, the Hamiltonian separate into a Hamiltonian on non-interacting fermions,
\begin{align}
\hat{\mathcal{H}}^\textrm{f} =&\ \sum_{m,{\bf k},\sigma} \left( Z_m\, \varepsilon^{(0)}_{m,\bf k} + \epsilon_m - \mu - \lambda_m \right) \hat{f}_{{\bf k}m\sigma}^\dag \hat{f}_{{\bf k}m\sigma}\ , \label{fermionic-hamiltonian}
\end{align}
and a purely slave-spin single-site Hamiltonian $\hat{\mathcal{H}}^\textrm{s}=\hat{\mathcal{H}}^\textrm{s}_0+\hat{\mathcal{H}}^\textrm{s}_U+\hat{\mathcal{H}}^\textrm{s}_J$, with
\begin{align}
\hat{\mathcal{H}}^\textrm{s}_0 =&\ \sum_{m,\sigma} \left( h_m^*\, \hat{O}_{m \sigma} + h_m\, \hat{O}^\dag_{m\sigma} \right)
+  \sum_m \lambda_m \sum_\sigma \left( \hat{S}^z_{m\sigma}+\frac{1}{2} \right) , \label{ss-hamiltonian-0}\\
\hat{\mathcal{H}}^\textrm{s}_U =&\ U \sum_m \hat{S}^z_{m\uparrow} \hat{S}^z_{m\downarrow} + U' \sum_{m\neq m'} \hat{S}^z_{m\uparrow} \hat{S}^z_{m'\downarrow} \nonumber \\
&\ + \left( U' - J \right) \sum_{m< m',\sigma} \hat{S}^z_{m\sigma} \hat{S}^z_{m'\sigma}\ , \label{ss-hamiltonian-U}\\
\hat{\mathcal{H}}^\textrm{s}_J =&\ - J_X \sum_{m\neq m'} \hat{S}^+_{m\uparrow} \hat{S}^-_{m\downarrow}  \hat{S}^+_{m' \downarrow} \hat{S}^-_{m' \uparrow} \nonumber \\
&\ + J_P \sum_{m\neq m'}   \hat{S}^+_{m\uparrow} \hat{S}^+_{m\downarrow}  \hat{S}^-_{m' \downarrow} \hat{S}^-_{m' \uparrow}\ , \label{ss-hamiltonian-J}
\end{align}
where $Z_m = \langle  \hat{O}^\dag_{m\sigma} \rangle\, \langle  \hat{O}_{m\sigma} \rangle$ is the hopping renormalisation factor and the quasiparticle weight for the orbital $m$, and $h_m=\langle  \hat{O}^\dag_{m\sigma} \rangle\, \overline{\varepsilon}^{(m)}_{(T)}$, with
\begin{align}
\overline{\varepsilon}^{(m)}_{(T)} =&\ \sum_{j (\neq i) }  t^{m}_{ij} \langle \hat{f}^\dag_{im\sigma} \hat{f}_{jm\sigma} \rangle = \frac{1}{N_s} \sum_{\bf k} \varepsilon^{(0)}_{m,\bf k}\, \langle \hat{f}_{{\bf k}m\sigma}^\dag \hat{f}_{{\bf k}m\sigma} \rangle \nonumber \\
=&\  \int_{-\infty}^\infty \varepsilon\, \rho_m (\varepsilon)\, n_m (\varepsilon) \textrm d \varepsilon \label{energy-uncorr}
\end{align}
the average electronic kinetic energy of the band $m$.
Here $\varepsilon^{(0)}_{m,\bf k} = \sum_{j(\neq i)} t^{m}_{ij}\, e^{-i {\bf k}\cdot ({\bf R}_j-{\bf R}_i)}$ is the dispersion relation of the uncorrelated $m$-orbital, $\rho_m (\varepsilon)$ its bare density of states, $n_m (\varepsilon)=\left( 1+e^{\beta (Z_m \varepsilon  + \epsilon_m - \mu - \lambda_m)} \right)^{-1}$ is the Fermi function for the orbital occupation per spin, and $\hat{f}_{im\sigma}=\frac{1}{\sqrt{N_s}} \sum_{\bf k} e^{-\imath {\bf k}\cdot {\bf R}_i} \hat{f}_{{\bf k}m\sigma}$ is the Fourier transform to the reciprocal space for the fermionic operators.
For convenience, we take the parameter $c_{im\sigma}$ in $\hat{O}_{im\sigma}$ to be real, hence $\langle \hat{O}_{m\sigma} \rangle=\sqrt{Z_m}$ is real too, and $h_m=\sqrt{Z_m}\overline{\varepsilon}^{(m)}_{(T)}$.\cite{Hassan2010,DeMedici2017a}

The total Hamiltonian, without taking into account the $\mu$ and energy shifts, is $\hat{\mathcal{H}}=\hat{\mathcal{H}}^\textrm{f}+N_s\, \hat{\mathcal{H}}^\textrm{s}-2\, N_s\, E_{MF}$, where
\begin{align}
E_{MF} =&\ \sum_{\substack{j (\neq i) \\ m,m',\sigma}}  t^{mm'}_{ij} \langle \hat{O}^\dag_{im\sigma} \rangle \langle \hat{O}_{jm' \sigma} \rangle \langle \hat{f}^\dag_{im\sigma} \hat{f}_{jm' \sigma} \rangle \nonumber \\
=&\ 2 \sum_m Z_m \overline{\varepsilon}^{(m)}_{(T)}
\end{align}
is the energy per site of the hopping Hamiltonian at the mean-field level.
The free energy density is $f = -\frac{1}{\beta} \ln \mathcal{Z}_\textrm{i}$, with $\mathcal{Z}_\textrm{i} =  \textrm{Tr} \left( e^{-\beta\, \frac{\hat{\mathcal{H}}}{N_s}}\right)$ the partition function for one site,
\begin{align}
f =&\ - \frac{2}{\beta} \sum_m \int_{-\infty}^{\infty} \rho_m(\varepsilon)\, \ln \left( 1+e^{-\beta \left(Z_m \varepsilon -\mu -\lambda_m \right)}\right) \textrm{d}\varepsilon \nonumber \\
   &\ - \frac{1}{\beta} \ln \left( \mathcal{Z}_1^s \right) \label{free-energy} \\
   &\ - 4\, \sum_m Z_m\, \overline{\varepsilon}^{(m)}_{(T)} \nonumber\ .
\end{align}
The first line of Eq. \eqref{free-energy} is due to the fermionic degrees of freedom, the middle one is the slave-spin part with $\mathcal{Z}_1^s = \textrm{Tr} \left( e^{-\beta\, \hat{\mathcal{H}^\textrm{s}}}\right)_s$ its one-site partition function, and the bottom term is the mean-field energy of the hopping term.
By minimising $f$ against the mean-field parameters $Z_m$ and $\lambda_m$, we obtain the self-consistent equations,
\begin{align}
Z_m =&\ \frac{1}{4\, \overline{\varepsilon}^{(m)}_{(T)}}\, \left<\, h_m\, \left( \hat{O}_{m\uparrow} + \hat{O}_{m\downarrow} + \hat{O}^\dag_{m\uparrow} + \hat{O}^\dag_{m\downarrow} \right)\, \right>_s\ , \label{self-1} \\
n_m =&\ \left<\, \left( \hat{S}^z_{m\uparrow} + \hat{S}^z_{m\downarrow} + 1 \right)\, \right>_s\ , \label{self-2}
\end{align}
where $\langle \hat{A} \rangle_s = \frac{1}{\mathcal{Z}_1^s}\textrm{Tr} \left( \hat{A}\, e^{-\beta \hat{\mathcal{H}}^s}\right)_s$ is the expectation value calculated on the slave-spin Hilbert space, and is solved by diagonalising the $16\times16$ matrix for the single-site Hamiltonian $\hat{\mathcal{H}}^s$.

In the Appendix \ref{opc-append} we calculate the finite temperature extension of the real choice of the $c$-parameter for one band.
We obtain a self-consistent equation $\mathcal{F}(c, T, \overline{\varepsilon}_0, n_0)=0$ that depends on the $c$-parameter, the temperature $T$, the uncorrelated kinetic energy for one band, i.e., $\overline{\varepsilon}_0=\overline{\varepsilon}_{(T)}\vert_{Z=1,\lambda=0}$, and the occupation number of the band in the non-interacting limit $n_0$.
The SSMF yields a non-zero  Lagrange multiplier in the uncorrelated limit, $\lambda_0$,\cite{DeMedici2014,DeMedici2017,DeMedici2017a}
an unwanted behaviour that is solved by shifting $\lambda$ to satisfy the physical non-interacting limit $\lambda=0$.
Previous works use a numerical calculation of $\lambda_0$ to perform this shift.
In our case, we obtain in Eq. \eqref{lambda0} an analytic expression for $\lambda_0$ that depends on $c$, $ \overline{\varepsilon}_0$ and $n_0$, and use it throughout our calculations.

In the multiorbital case, the non-interacting limit is just a set of uncoupled one-band systems, and each orbital $m$ has its $c_m$ and $\lambda_{m,0}$ determined by $T$, $\overline{\varepsilon}^{(m)}_0$, and its non-interacting occupation number $n_{m,0}$.
For a fixed temperature $T$ and total occupation of the site $n_\textrm{site}$, the iteration scheme used is: (i) calculate the non-interacting chemical potential through $n_\textrm{site}=\sum_m n_{m,0}$, and consequently each orbital occupation $n_{m,0}$ and kinetic energy $\overline{\varepsilon}^{(m)}_0$ (ii) set the value of $c_m$ using the self-consistent equation, and consequently the value of $\lambda_{m,0}$, and (iii) solve the self-consistent Eqs. \eqref{self-1} and \eqref{self-2} for $Z_m$ and $\lambda_m$.

\section{Results}
\label{sec:results}

\subsection{Isotropic case (\texorpdfstring{$W_1=W_2$}{W1 = W2})}

\subsubsection{Quasiparticle weight and coherence temperature}

\begin{figure}
  \includegraphics*[width=0.46\textwidth]{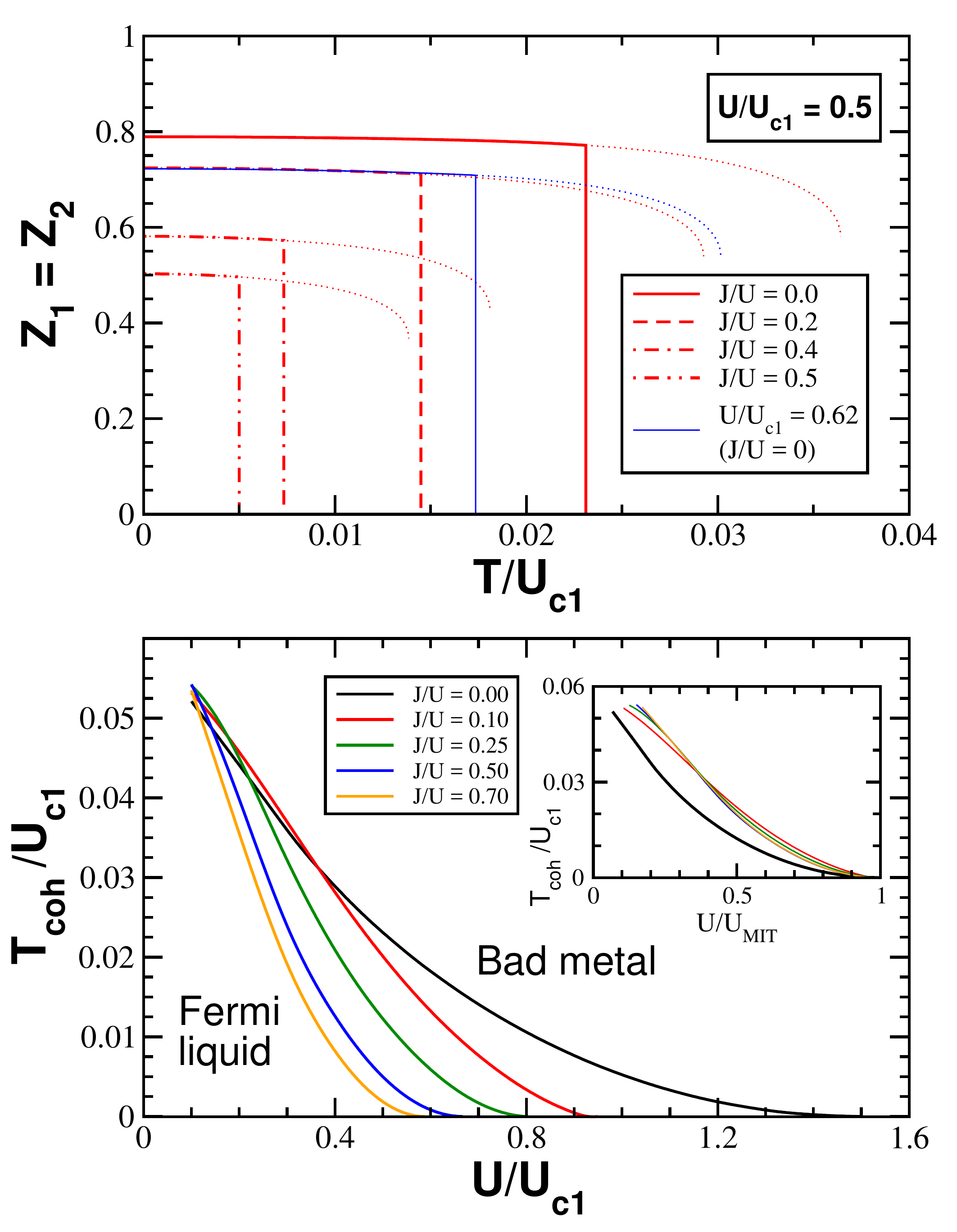}
  \caption{ \label{fig:n100-Z-Tcoh} Destruction of the Fermi liquid by Hund's coupling.
  {\bf Top: }Quasiparticle weight for half-filling as a function of temperature, with $U/U_\textrm{c1}=0.5$ (red) for different $J/U$.
  See Appendix \ref{solutions-append} for details in the construction of the physical solution.
  An increase in temperature from $T=0$ slightly reduces the quasiparticle weight $Z_m$ on each band, and at $T=T_\textrm{coh}$ a first-order transition to the trivial state with $Z=0$ occurs.
  We identify the trivial solution with a bad metal state, and $T_\textrm{coh}$ with the coherence temperature associated with the crossover to the bad metal regime.
  Also in thin solid blue line, we show the case $U/U_\textrm{c1}=0.62$ and $J=0$ that is comparable to the $U/U_\textrm{c1}=0.5$ and $J/U=0.2$ one.
  {\bf Bottom: }Phase diagram for temperature versus interaction.
  We plot the coherence temperature $T_\textrm{coh}$ as function of the interaction $U$, for different values of the Hund's coupling $J/U = 0.0$, $0.1$, $0.25$ and $0.5$.
  An increase in Hund's coupling $J$ increases the correlations, driving the system closer to a MIT, significantly decreasing $T_\textrm{coh}$, and enhancing the stability of the bad metal state.
  {\bf Inset:} Same plot with the interaction $U$ normalised with the value at the metal insulator transition $U_\textrm{MIT}$.
  The universal behaviour of $T_\textrm{coh}$ when varying $J$ reflects that changes in the Hund's rule interaction affect the system only through $U_\textrm{MIT}$.}
\end{figure}

For fixed $U$, we show how increasing $J$ drives the metal close to the Mott insulating phase and reduces the coherence temperature.
In the top of Fig. \ref{fig:n100-Z-Tcoh}
we plot in red the quasiparticle weight $Z_1=Z_2=Z$ for half-filling, $U/U_\textrm{c1}=0.5$ and vary $J/U$.
In thin dotted lines we plot the solutions with finite $Z$ in the whole range of temperature where they are found to exist. These solutions are the ones that evolve continually from the zero-$T$ results.
For each case, at $T=T_\textrm{coh}$ the free energy of this finite-$Z$ solution crosses with the one corresponding to the ``\emph{trivial}'' solution  with $Z_1=Z_2=0$, and the later becomes the stable phase for $T>T_\textrm{coh}$.
For a description of the construction of the physical solution refer to Appendix \ref{solutions-append}.
For $T<T_\textrm{coh}$ the quasiparticle weight is almost constant, slightly decreasing with increasing $T$, and jumping to zero at $T=T_\textrm{coh}$.
We identify this temperature with the coherence temperature of the fermionic quasiparticles, associated with the crossover to a bad metallic state.

It is known that the effect of the Hund's coupling in multiorbital systems at half-filling is to increase correlations, and increasing $U$ and/or $J$ reduce the values of $Z$ and $T_\textrm{coh}$,\cite{DeMedici2011a,Georges2013a} enhancing the stability of the bad metal.
But the reduction in the coherence temperature is more pronounced than the reduction in the quasiparticle weight, showing that the reduction of $T_\textrm{coh}$ is not just due to band renormalisation.\cite{Mezio2017}
We also add in blue the solution for $J=0$ and $U/U_\textrm{c1}=0.62$, that has the same quasiparticle weight at zero-$T$ as the $U/U_\textrm{c1}=0.5$ and $J/U=0.2$ case.
A similar increase in correlations at $T=0$, due to $J$ or $U$, gives a slightly different reduction of $T_\textrm{coh}$.
This difference decreases when correlations are stronger (bigger $J$ and/or $U$) and the system gets closer to a MIT, and we can conclude that regarding $T_\textrm{coh}$ there is no big difference here between the increase of correlations through $J$ or $U$.
Although, there is a qualitative difference between the $J=0$ and $J>0$ case, that we discuss below.

In the bottom of Fig. \ref{fig:n100-Z-Tcoh} we plot the phase diagram for temperature $T$ vs interaction $U$ and how it changes with increasing $J/U$. The solid lines are the coherence temperature $T_\textrm{coh}$ associated with the crossover to a bad metal regime. We can see that the SSMF method obtains a very low coherence temperature close to the Mott MIT, as expected for strongly correlated materials.
The inset of Fig. \ref{fig:n100-Z-Tcoh} shows the same plot, but with $U$ normalised to its value at the metal insulator transition, $U_\textrm{MIT}$.
We can see that the $J=0$ case is qualitatively different from those with finite $J$, whereas for the latter $T_\textrm{coh}$ becomes independent of the particular value of $J/U>0$, and only depends on the relative position of $U$ to its zero-$T$ metal insulator transition value.
In other words, for two equal bands $T_\textrm{coh}$ only depends on $J/U$ through its effect on the zero temperature critical interaction $U_\textrm{MIT}$.

It is useful to focus on the atomic configurations to understand the difference between zero- and finite-$J$.
A detailed list of the atomic states and its energies is in Appendix \ref{atomic-append}.
At half-filling, it is easy to check from Eqs. \eqref{HU} and \eqref{HJ} that the states with two electrons onsite have the lowest atomic energy.\cite{Koga2004,DeMedici2011,Komijani2017}
For $J=0$, this low atomic energy sector is six-fold degenerate with energy $E=-2\, U$, and has states with total spin per site $S=0$ and $S=1$.
While for $J>0$, this degeneracy is lifted into the two different spin sectors, each of them now three-fold degenerate.
Here, the triplet $S=1$ sector has the lowest atomic energy, $E_{S=1}<E_{S=0}$.
This change in the degeneracy and total spin of the lowest energy manifold is the causes of the different behaviour of the ground state for $J=0$ and $J>0$,\cite{Koga2004,DeMedici2017a} and consequently, cause the different behaviour observed in this work at low temperature.

The sector next in energy is $E_{S=\frac{1}{2}}$, which have one/three electrons per site and is eight-fold degenerate.
For $J/U<\frac{1}{3}$ we have that $E_{S=0}<E_{S=\frac{1}{2}}$, but this inequality is reversed when $J/U>\frac{1}{3}$ (see Fig. \ref{fig:energies-hf}) and we can expect a change in the finite temperature behaviour.
We find no qualitative change when $J/U$ crosses this value and, as discussed before, the quasiparticle weight $Z$ has an almost constant value for $T<T_\textrm{coh}$.
This tells us that the effective temperature is much lower than any energy scale of the system and, within this range of temperatures the system lives mostly in the low energy manifold.

\subsubsection{Charge and spin fluctuations}

\begin{figure}
  \includegraphics*[width=0.46\textwidth]{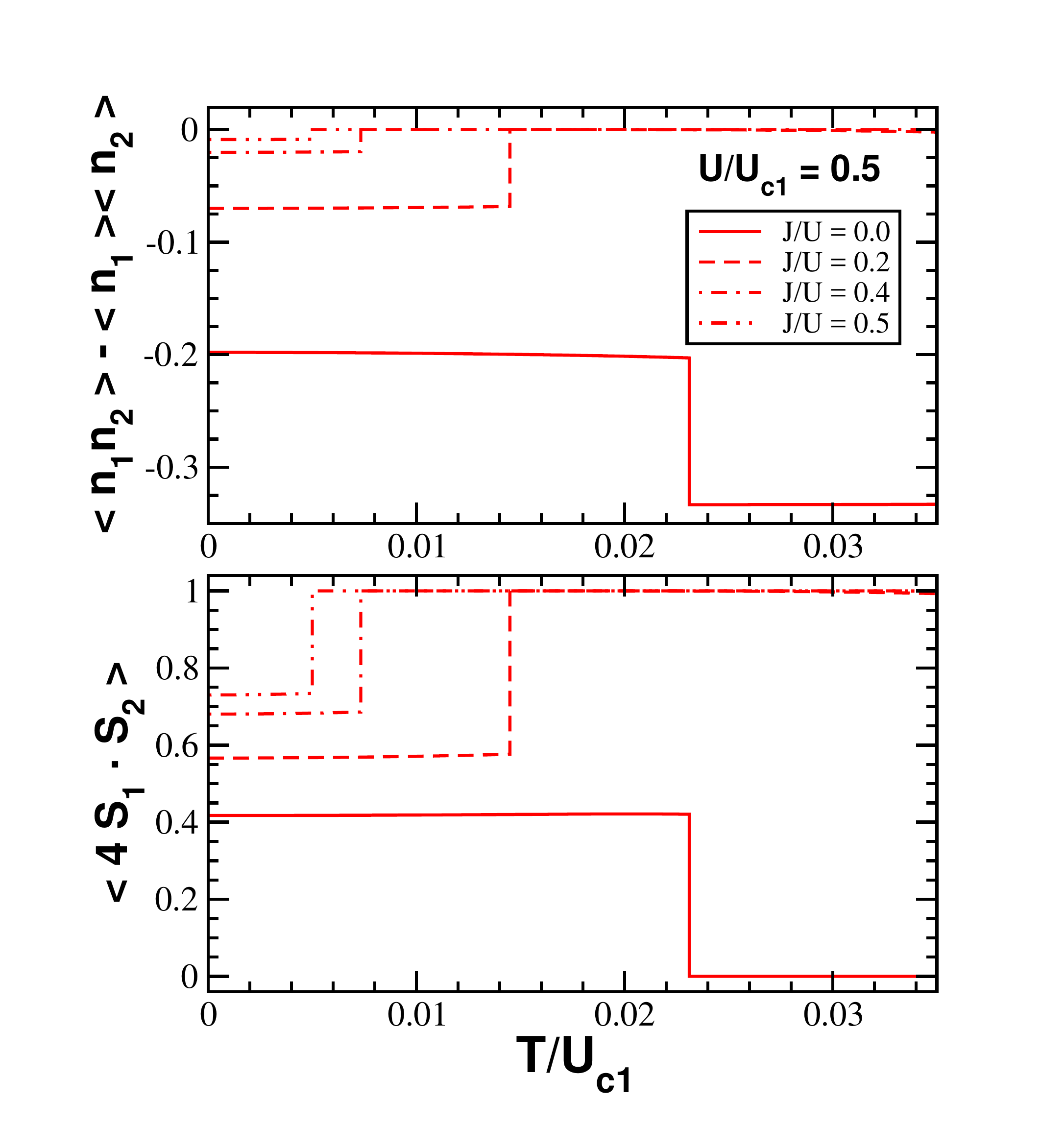}
  \caption{\label{fig:n100-correls} {\bf Top:} Inter-orbital charge correlations for half-filling with $U/U_\textrm{c1}=0.5$, varying $J/U$.
For $T<T_\textrm{coh}$, increasing $J$ increases localisation, the system prefers to have one electron per orbital, and the charge fluctuation approaches to zero.
At the transition $T=T_\textrm{coh}$ the charge fluctuations jump to zero for $J>0$.
For $J=0$ it jumps to $-1/3$, which is the same limit value when $U\rightarrow U_\textrm{MIT}$ (see main text). High temperature limit for all the cases is zero (not shown).
{\bf Bottom:} Inter-orbital physical spin fluctuations $\langle 4\, \hat{\vec{\mathbb{S}}}_1 \cdot \hat{\vec{\mathbb{S}}}_2 \rangle$, same parameters as above.
For $T<T_\textrm{coh}$, increasing $J$ increases electron localisation and polarise their spins (one electron per orbital and fully parallel spins).
The limit value $\langle 4\, \hat{\vec{\mathbb{S}}}_1 \cdot \hat{\vec{\mathbb{S}}}_2 \rangle=1$ is achieved at the transition temperature $T=T_\textrm{coh}$.
For $J=0$ there is no inter-orbital effective interaction and the correlation between orbital spins is zero in the trivial state, at $T>T_\textrm{coh}$.
High temperature limit for all the cases is zero (not shown).}
\end{figure}

The calculation of spin and charge fluctuations is helpful to clarify the different nature of the correlations, either due to an increase of $J$ or $U$.\cite{Fanfarillo2015}
We remember here that the occupation operator of the orbital $m$ is $\hat{n}_m=\hat{n}_{m\uparrow}+\hat{n}_{m\downarrow}$ and the $z$-component of its physical spin is $2\, \hat{\mathbb{S}}_m^z=\hat{n}_{m\uparrow}-\hat{n}_{m\downarrow}$, and we use the constraint Eq. \eqref{constraint1} to write them in the slave-spin Hilbert space.
The transversal components of the physical spin $\hat{\mathbb{S}}_m^{\pm}$ are also easy to express in the local slave-spin basis.
In this way, in Fig. \ref{fig:n100-correls}
we plot the on-site inter-orbital charge (top) and spin (bottom) fluctuations, $\langle \hat{n}_1\, \hat{n}_2\rangle_s - \langle \hat{n}_1\rangle_s\, \langle\hat{n}_2\rangle_s$ and $\langle 4\, \hat{\mathbb{\bf S}}_1 \cdot \hat{\mathbb{\bf S}}_2 \rangle_s$, respectively.
Here $\hat{\mathbb{\bf S}}_m$ stands for the spin operator of the orbital $m$ on the site, $(\hat{\mathbb{S}}_m^x,\hat{\mathbb{S}}_m^y,\hat{\mathbb{S}}_m^z)$.

Regarding the inter-orbital charge fluctuations (top of Fig. \ref{fig:n100-correls}), they are negative for $T<T_\textrm{coh}$, as expected for repulsive interactions in a FL.
As it is known from Mott physics, an increase of correlations localise electrons on sites and suppresses charge fluctuations.
But again, the effects are different whether we increase correlations through $U$ or $J$.\cite{Fanfarillo2015}
The increase of the Hund's coupling polarizes the spin on the site and increase the energy gap between $E_{S=1}$ and $E_{S=0}$, restricting the system to the $S=1$ triplet configurations, and making the inter-orbital charge fluctuations approach to zero.
At the transition $T=T_\textrm{coh}$ a jump to zero (for $J>0$) at the trivial solution occurs, where this state can be understood within the method as two flat bands interacting each other through an effective ferromagnetic interaction.
On the other side, for $J=0$, increasing correlations via the Coulomb interaction $U$ makes the inter-orbital charge fluctuations of the FL phase approach to the value $-1/3$ (not shown in plot).
We can understand this by noting that, if we restrict ourselves to the six-fold degenerate manifold, two states have $\langle \phi \vert\, \hat{n}_1\, \hat{n}_2\, \vert  \phi \rangle = 0$ and the other four $\langle \phi \vert\, \hat{n}_1\, \hat{n}_2\, \vert  \phi \rangle = 1$, obtaining a total $\langle  \hat{n}_1\, \hat{n}_2 \rangle= 2/3$ that account for the value $-1/3$ in the charge fluctuation quantity (see Appendix \ref{atomic-append}).
The next available states are at an energy gap of $E_{S=\frac{1}{2}}-E_{S=0,1}=\frac{U}{2}$, and are accessible to the system through the hopping Hamiltonian.
Finally, increasing $U$ towards the MIT makes this gap bigger and the system more restricted to the lowest energy manifold, where we shown that $\langle \hat{n}_1\, \hat{n}_2 \rangle -\langle \hat{n}_1\rangle\langle \hat{n}_2 \rangle = -\frac{1}{3}$.
We can interpret the jump to $-1/3$ in the $J=0$ line in a similar way: the transition at $T=T_\textrm{coh}$ to the trivial phase with $Z=0$ cancels the effect of the hopping Hamiltonian, restricting the system to the lowest manifold and obtaining the value $-1/3$ at low temperatures.
When increased further the temperature, at around $T\sim U/2$ thermal transitions to other available states occur, making the inter-orbital charge fluctuation to approach slowly to zero (not shown in plot).

The spin fluctuations (bottom of Fig. \ref{fig:n100-correls}) grow with increasing $J/U$, as expected for the spin polarisation due to Hund's coupling.
The value $1$ is the correct limit for the picture of a ground state lying in the $E_{S=1}$ manifold.
Similar arguments as before apply for the transition to the trivial state for $J>0$, where two flat bands are coupled with $J$ and have a low temperature value $\langle 4\, \hat{\vec{\mathbb{S}}}_1 \cdot \hat{\vec{\mathbb{S}}}_2 \rangle=1$.
For $J=0$ in the trivial state there are no interactions between the spins and $\langle 4\, \hat{\vec{\mathbb{S}}}_1 \cdot \hat{\vec{\mathbb{S}}}_2 \rangle=0$.
We can say that for $T>T_\textrm{coh}$ there is a qualitative difference in spin-triplet correlation between $J=0$ and $J\neq 0$.
Finally, further increase in the temperature washes out the effect of $J$ on the trivial states, and at high temperatures we approach the limit $\langle 4\, \hat{\vec{\mathbb{S}}}_1 \cdot \hat{\vec{\mathbb{S}}}_2 \rangle=0$ (not shown in plot).

\subsection{Anisotropic orbitals}

We explore now the effect of the Hund's coupling at finite temperature when we have an orbital anisotropy, $W_2/W_1<1$, i.e., the two bands have different widths. We will see how an orbital-selective bad metal becomes possible.

\subsubsection{Quasiparticle weights, coherence temperatures and fluctuations}

\begin{figure}
  \includegraphics*[width=0.46\textwidth]{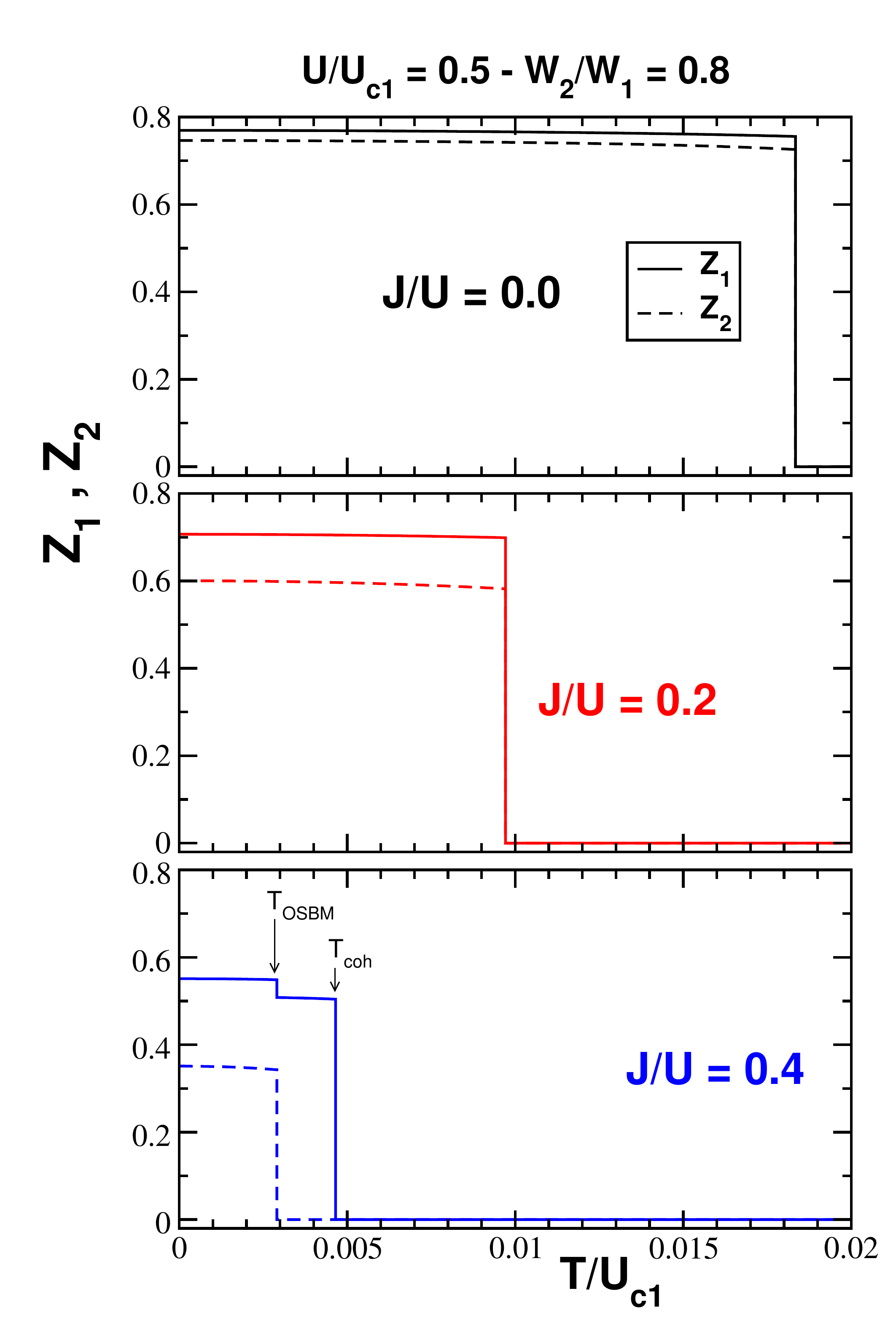}
  \caption{\label{fig:aniso-Z} Hund's rule coupling produces an orbital-selective bad metal.
  Quasiparticle weights $Z_1$ and $Z_2$ for half-filling and $W_2/W_1 = 0.8$, with $U/U_\textrm{c1}=0.5$.
  From top to bottom, $J/U=0,\ 0.2$ and $0.4$.
  For $J/U = 0.4$ an intermediate transition to an orbital-selective bad metal (OSBM) occurs at $T_\textrm{OSBM}$, where $Z_2$ vanishes while $Z_1$ is further renormalised.
  After this transition the inter-orbital charge fluctuations vanish (c.f. Figure \ref{fig:aniso-correls} top) while spin fluctuations increase (c.f. Figure \ref{fig:aniso-correls} bottom), in a similar manner to the first-order transition in the isotropic case.}
\end{figure}

In Figure \ref{fig:aniso-Z}
 we show the quasiparticle weight for orbital $1$ (solid lines) and $2$ (dashed lines), where the bandwidth anisotropy is $W_2/W_1=0.8$ and $U/U_\textrm{c1}=0.5$, for three values of Hund's coupling, $J/U=0.0$, $0.2$ and $0.4$.
Same as before, the dotted lines show the different solutions in the whole temperature range they are found, and we set the different first-order transitions where the different free energies crosses each other (see Appendix \ref{solutions-append}).
The increase of correlations affects more to the narrow orbital, as we can see from the stronger renormalisation of $Z_2$ when $J$ increases.
Although the orbital anisotropy is mild, with $J/U=0.4$ we find an intermediate transition to a solution with $Z_1>0$ and $Z_2=0$.
Following our previous interpretation, here the narrow orbital $2$ transitions to a bad metal state while the wide orbital $1$ remains FL, although it suffers a renormalisation in the quasiparticle weight $Z_1$.
We call this phase an \emph{orbital-selective bad metal} (OSBM), and the transition temperature $T_\textrm{OSBM}$.
Further increasing temperature, another first-order transition occurs at $T=T_\textrm{coh}$, where the wide orbital $1$ also collapse and both bands are in a bad metal state.
For this value of the correlation $U$, an increase of $J$ stabilizes the OSBM phase.

\begin{figure}
  \includegraphics*[width=0.46\textwidth]{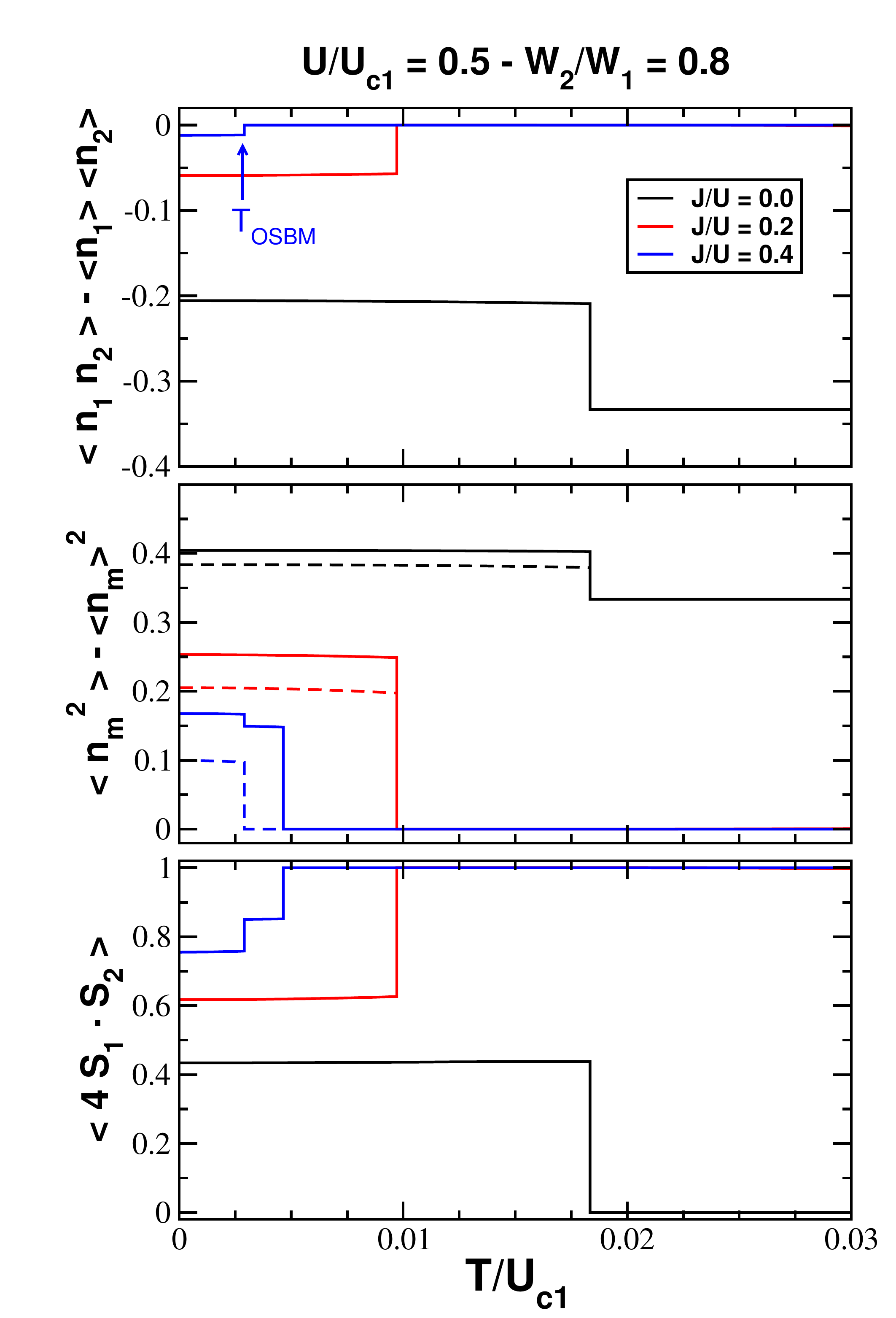}
  \caption{ \label{fig:aniso-correls} Charge and spin fluctuations for $n=1$, with $W_2/W_1 = 0.8$ and $U/U_\textrm{c1}=0.5$.
  Black, red and blue are for $J/U=0$, $0.2$ and $0.4$, respectively.
  {\bf Top:} Inter-orbital charge correlations. Same as in the isotropic case, increasing $J$ with $T<(T_\textrm{coh},T_\textrm{OSBM})$ increases electronic localisation, and the charge fluctuation approaches to zero.
  In the OSBM phase there are no charge fluctuation between orbitals ($J/U=0.4$ for $T>T_\textrm{OSBM}$).
  {\bf Middle:} Intra-orbital charge fluctuations.
  Solid (dashed) line refers to the wide (narrow) orbital.
  In the OSBM phase the wide orbital has local charge fluctuations, while no fluctuations of the charge in the other orbital occurs.
  This means that in the OSBM phase the wide orbital remains a metal with renormalised quasiparticles, while the narrow orbital is fully localised.
  {\bf Bottom:} Inter-orbital spin fluctuations. An increase in spin fluctuations occurs when transitioning to the OSBM phase.}
\end{figure}

The physics of this new phase is better understood from the charge and spin fluctuations, which we show in Figure \ref{fig:aniso-correls}.
From top to bottom we have inter-orbital charge, intra-orbital charge, and inter-orbital spin fluctuations.
In black, red and blue we show result for $J/U=0.0$, $0.2$ and $0.4$, respectively.
Solid and dashed lines in the middle plot refers to wide and narrow orbitals, respectively.
For the inter-orbital charge fluctuations (top) we have the same general behaviour as in the isotropic case of the previous section.
The new aspect is that for $J/U=0.4$ the jump to zero occurs at the transition to the OSBM phase, at $T_\textrm{OSBM}$.
This means that charge movement between orbitals cancels when the narrow orbital collapses.
The narrow orbital is completely localised in this phase, and interaction with the FL of the wide orbital is only through the spins and the Hund's rule coupling.
The intra-orbital charge fluctuations $\langle \hat{n}_m^2\rangle - \langle\hat{n}_m\rangle^2$ (middle) shows clearly that when $T_\textrm{OSBM}<T<T_\textrm{coh}$  the wide band is still metallic (solid blue line) while the narrow band has $\langle \hat{n}_2^2\rangle = \langle\hat{n}_2\rangle^2$.
When the transition to the OSBM phase occurs, $Z_2=0$ and the system is more restricted to the $S=1$ triplet configuration, which explains the rise on the spin fluctuations (bottom).
The non-interacting limit ($U=J=0$) for these qunatities are: $0$ for the inter-orbital, $0.5$ for the intra-orbital, and $0.25$ for the inter-orbital spin fluctuations.

\subsubsection{Phase diagrams}

\begin{figure}
    \includegraphics[width=0.46\textwidth]{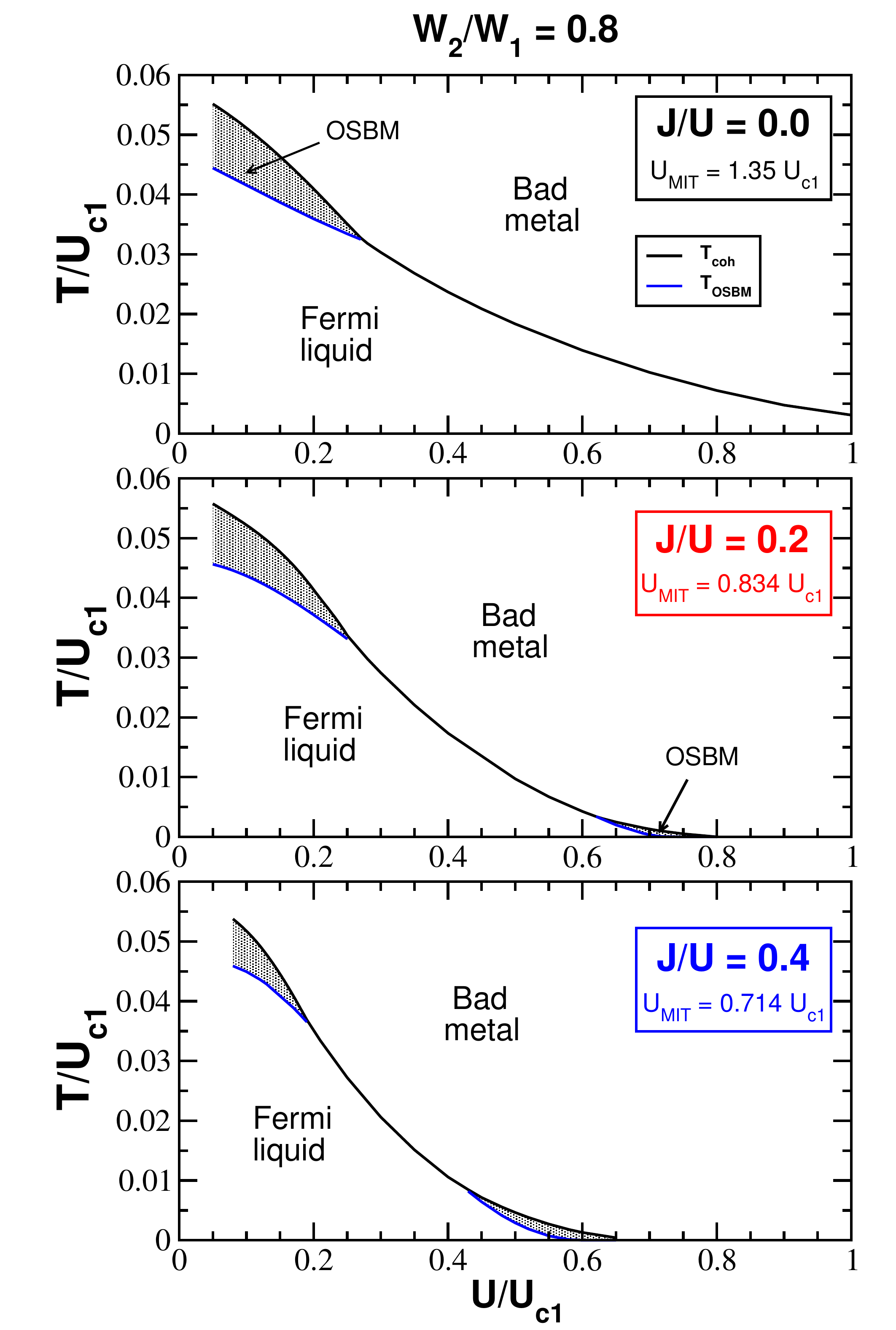}
  \caption{\label{fig:aniso-Tcoh-moveJ} Phase diagrams for $T$ vs $U/U_\textrm{MIT}$, for different values of Hund's coupling, with $W_2/W_1=0.8$.
  The transition temperatures $T_\textrm{OSBM}$ and $T_\textrm{coh}$ are in blue and black solid lines, respectively.
  For simplicity, OSBM phase regions are shaded grey.
  Red arrows mark $U=0.5\, U_{c1}$ on each case.
  In addition to the strong reduction of the critical interaction $U_\textrm{MIT}$ for finite $J$, Hund's coupling enhances the strong interaction OSBM region and diminishes the weak interaction OSBM region.}
\end{figure}

We now consider how the transition temperatures to the bad metal ($T_\textrm{coh}$) and the orbital-selective bad metal ($T_\textrm{OSBM}$) vary as a function of the interaction strenghts $U$ and $J$.
In Figure \ref{fig:aniso-Tcoh-moveJ}
we plot for $W_2/W_1=0.8$ and different values of $J/U$, the phase diagram for temperature $T$ vs interaction $U$.
The transition temperatures $T_\textrm{OSBM}$ and $T_\textrm{coh}$ are in blue and black, respectively, and the OSBM regions are shaded in grey.
The first thing to notice is the appearance of an OSBM region at weak $U$, which can be related to a weak inter-band coupling.
The increase of $J/U$ reduces even further the inter-band coupling ($U'$ and $U'-J$), and the low-$U$ OSBM region shrinks.
Also, a high-$U$ OSBM region appears for finite-$J$, which is the OSBM phase seen in Figs. \ref{fig:aniso-Z} and \ref{fig:aniso-correls} when $J/U=0.4$.
This region does not exist when $J=0$ and gets enhanced with increasing Hund's rule coupling.
We have now a more complete picture of how the Hund's rule coupling enhances the OSBM phase, as discussed in the previous section, seen from two main effects:
(i) the growth of the high-$U$ OSBM region with increasing $J$, and
(ii) the increase of correlations due to $J$ that strongly reduce $U_\textrm{MIT}$ and shift the OSBM region over the $U=0.5\, U_{c1}$ point.
Similar to the inset in Figure \ref{fig:n100-Z-Tcoh} there is a qualitative difference between $J=0$ and $J>0$ when $U$ is normalised with $U_\textrm{MIT}$ (not shown in the plot), i.e., the $T_\textrm{coh}$ vs $U/U_\textrm{MIT}$ lines for $J>0$ superpose each other and are different than the $J=0$ case.

\begin{figure}
  \includegraphics*[width=0.46\textwidth]{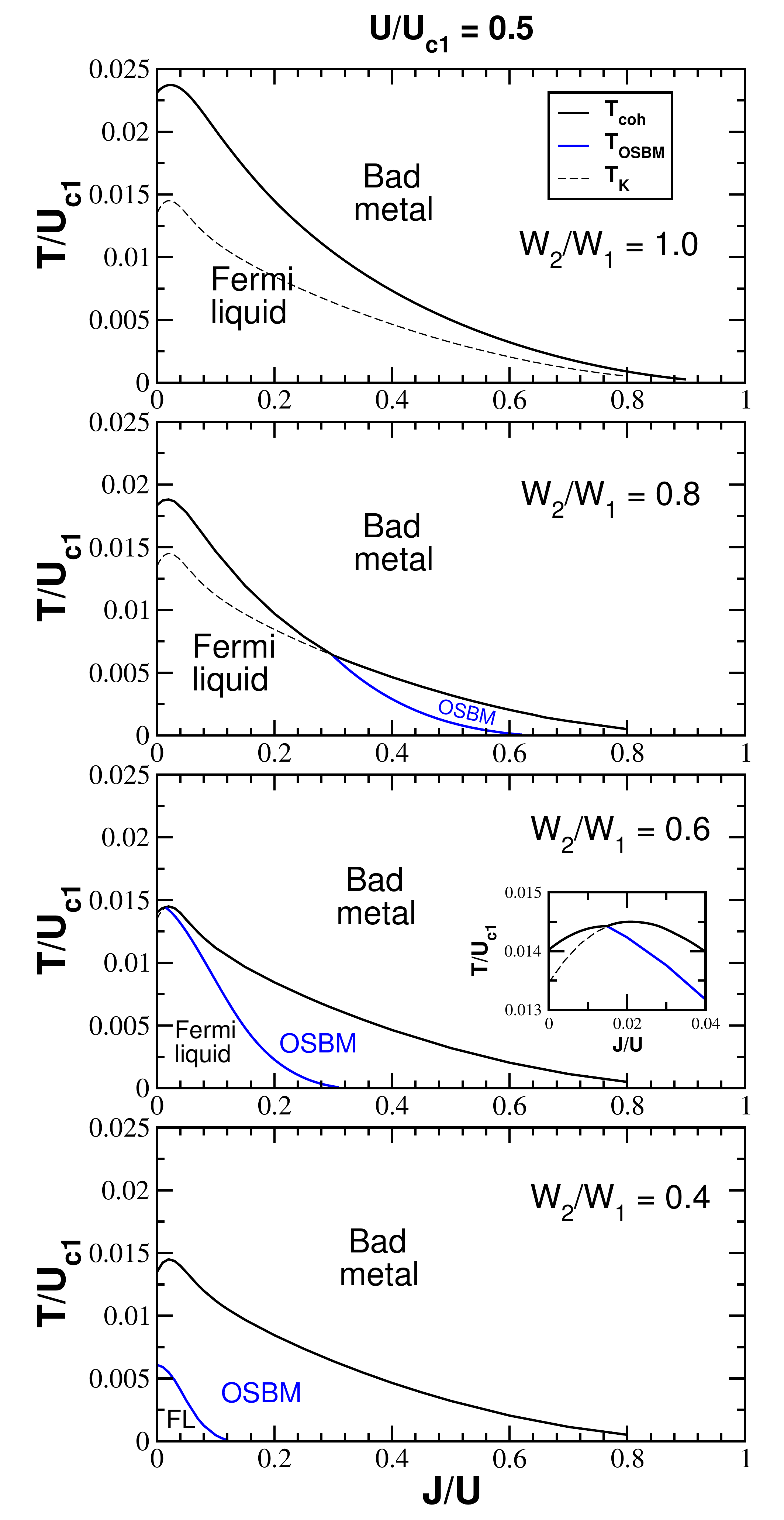}
  \caption{\label{fig:aniso-Tcoh-moveW} Stabilisation of the OSBM with increasing $J$ and $W_2/W_1$.
  Phase diagrams $T$ vs. $J$ for $U/U_\textrm{c1}=0.5$ and different anisotropies.
  For $W_2/W_1 < 1$ exist a critical $J_c$ where for $J>J_c$ and increasing temperature, the system goes a first-order transition at $T_\textrm{OSBM}$ (blue) toward an OSBM phase that is stable up to $T_\textrm{coh}$.
  At $T_\textrm{coh}$ (black) the quasiparticle weight of the wide band vanishes, and we have a bad metal in both bands.
  In the OSBM phase the narrow orbital is localised, in a high spin configuration, and no charge fluctuations between orbitals are present.
  The effective model of the system is a metallic band coupled to a localised band through only a spin-spin interaction, i.e., a ferromagnetic Kondo-Hubbard problem.
  This manifests in the fact that the $T_\textrm{coh}$ curves that are above in temperature to an OSBM phase become independent of the anisotropy, and has a decay that depends exponentially with the coupling $J$.
  We plot this temperature $T_K$ with a dashed line for $W_2/W_1= 1$, $0.8$ and $0.6$, while for $W_2/W_1= 0.4$ it coincide with $T_\textrm{coh}$ (black solid).}
\end{figure}

In Figure \ref{fig:aniso-Tcoh-moveW}
we plot for $U/U_{c1}=0.5$ and different values of $W_2/W_1$, the phase diagram for temperature $T$ vs Hund's coupling $J/U$.
The transition temperatures $T_\textrm{OSBM}$ and $T_\textrm{coh}$ are in blue and black, respectively.
The inset in $W_2/W_1=0.6$ is an enlargement of the low $J/U$ part.
An increase in anisotropy enhance the region where an OSBM phase exist, and an increase in $J/U$ favours this phase when anisotropy is present.
The $T_\textrm{coh}$ vs $J/U$ line when an OSBM phase is present at lower temperatures is always the same, disregarding the value of $W_2/W_1$.
When entering the OSBM region, the narrow orbital becomes flat ($Z_2=0$) and electron localise, while the wider orbital remains itinerant and interact with the full electron spin of the former through the Hund's coupling.
Considering the low energy physics of this phase the interaction $U'$ cancels, and the effective Hamiltonian for the system is a \emph{ferromagnetic Kondo lattice} with an additional Hubbard interaction ($U$) in the wide band.\cite{Biermann2005}
In this sense, the coherence temperature $T_\textrm{coh}$ is a Kondo temperature $T_\textrm{K}$ which does not depend on $W_2$ (dashed line in Fig. \ref{fig:aniso-Tcoh-moveW}).

\subsection{Entropy analysis}
\label{sec:entropy}

An interesting question is about how much of the temperature dependence of the entropy of the system can be captured by the SSMF method.
The slave-spin mapping \eqref{ss-map} increases the Hilbert space by the incorporation of a spin-$\frac{1}{2}$ degrees of freedom for every fermionic one, expanding it from a $16$-dimensional to a $16^2$-dimensional one.
The constraint \eqref{constraint1} removes any possible unphysical states, making the mapping exact.
But at the mean-field level, we impose the constraint only on average, allowing the participation of unphysical states and more specifically their contribution to the entropy.
\begin{figure}[ht]
  \includegraphics*[height=0.46\textwidth,angle=270]{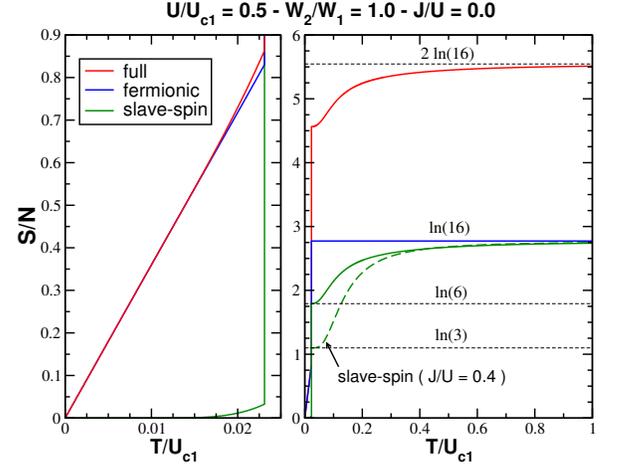}
  \caption{\label{fig:entro-iso} Entropy for the isotropic case with $J=0$.
  In red, blue and green we plot the total, fermionic and slave-spin entropy, respectively.
  {\bf Left:} For $T<T_\textrm{coh}$ the slave spin contribution is very small, and the fermionic degrees of freedom contribute almost all of the total entropy.
  {\bf Right:} For High-$T$, each of the $16$-dimensional subspaces (fermions/slave-spin) reach the corresponding value $\ln(16)$. The total entropy reflects the expansion of the Hilbert space to $16^2$ degrees of freedom per site and the approximation in the implementation of the constraint. Dashed green line is the slave-spin contribution to the entropy at the trivial solution for $J/U=0.4$.}
\end{figure}
\begin{figure}[ht]
  \includegraphics*[height=0.46\textwidth,angle=270]{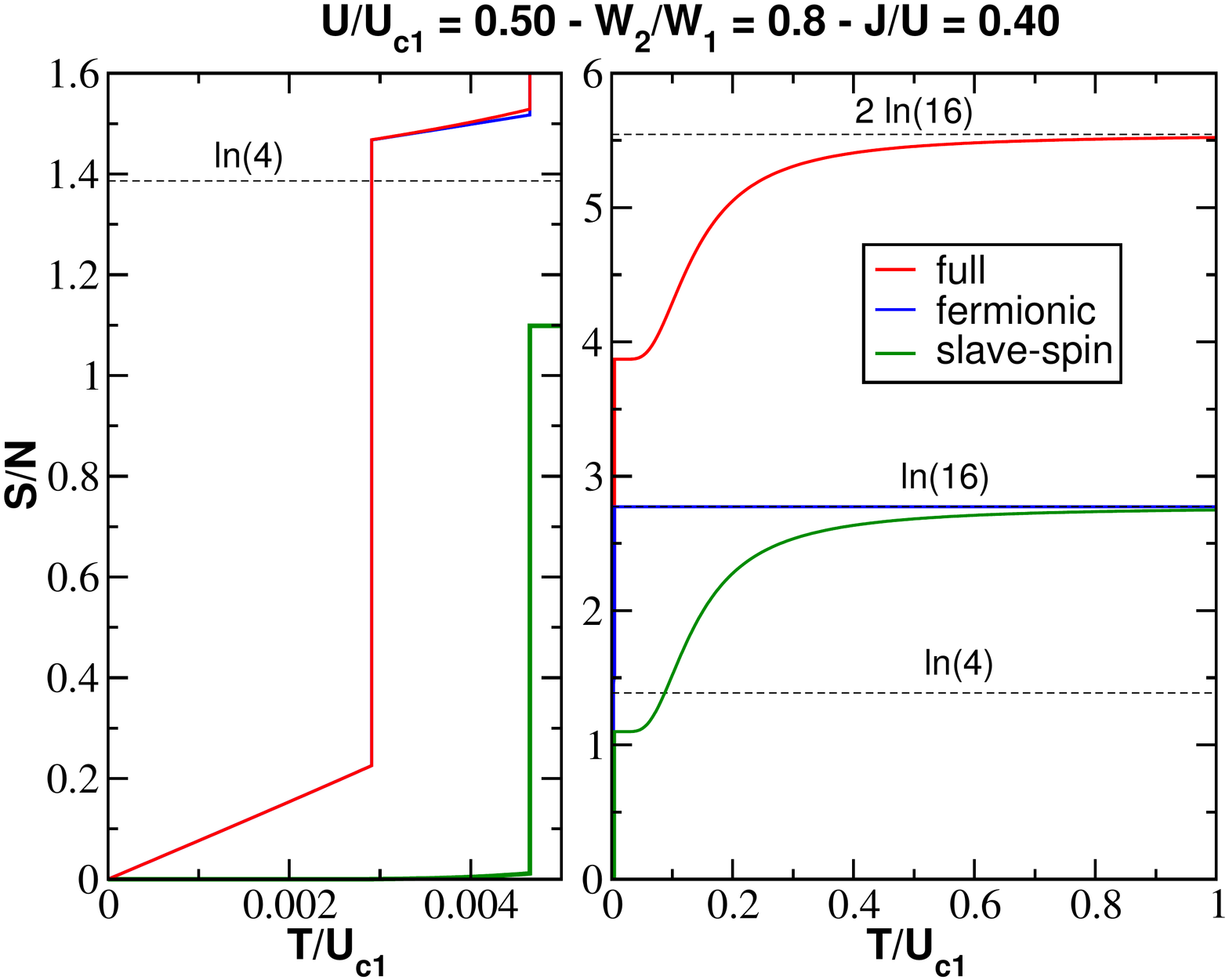}
  \caption{\label{fig:entro-aniso} Entropy for the anisotropic case where the OSBM phase occurs.
  In red, blue and green we plot the total, fermionic and slave-spin entropy, respectively.
  {\bf Left:} For $T<T_\textrm{coh}$ the slave spin contribution is very small, even when the OSBM phase occurs and $Z_2=0$. In this case, the fermionic degrees of freedom contribute almost all of the total entropy in the FL and OSBM phase.
  {\bf Right:} For High-$T$, each of the $16$-dimensional subspaces (fermions/slave-spin) reach the corresponding value $\ln(16)$. The total entropy reflects the expansion of the original Hilbert space to $16^2$ degrees of freedom per site and the approximated treatment of the constraint.}
\end{figure}
This is seen easily in the high temperature behaviour of the solid red lines in Figs. \ref{fig:entro-iso} and \ref{fig:entro-aniso}, where we plot the total entropy, incorporating all the degrees of freedom of the method.
The entropy (per site) is calculated using the thermodynamic relation
\begin{equation}
s(T)=\beta\, (\overline{u}-f)
\end{equation}
where $f$ is the free energy calculated in Eq. \eqref{free-energy}, and $\overline{u} = \frac{1}{\mathcal{Z}_\textrm{i}} \textrm{Tr} \left( \frac{\hat{\mathcal{H}}}{N_s}\, e^{-\beta\, \frac{\hat{\mathcal{H}}}{N_s}}\right)$ is the internal energy.
The decoupling of the fermionic degrees of freedom from the slave-spin ones at the mean-field level allows us to separate their contribution explicitly.
In the free energy, the first, second and third lines of Eq. \eqref{free-energy} are due to the fermions, the slave-spins and the hopping mean-field energy ($E_\textrm{NF}$), respectively.
The same occurs for $\overline{u}$, and we can separate the entropy in the fermionic and slave-spin contributions, $s=s_\textrm{f}+s_\textrm{ss}$, which are the blue and green lines, respectively, in Figs. \ref{fig:entro-iso} and \ref{fig:entro-aniso}.

Solid lines in Fig. \ref{fig:entro-iso} shows the isotropic case with $W_2=W_1$ and no Hund's coupling.
In the FL phase (left), for $T<T_\textrm{coh}$, we can see that the slave-spin contribution (green) is very small, which supports the atomic picture that the energy gap between the ground state and the low energy excitations is larger than the temperature scale in this range.
The linear behaviour of the total entropy (red) in the FL phase is all due to the fermionic contribution (blue), whose physics is that of the free electrons in a renormalised band.
For $T>T_\textrm{coh}$ we are in the trivial phase with $Z=0$.
Looking at the fermionic part of the free energy, we can think of $Z$ as a renormalisation factor for the inverse temperature $\beta=\frac{1}{T}$, and the trivial phase as the free fermion gas being in the infinite temperature limit with an entropy of $s_\textrm{f}=\ln (16)$.
For the slave-spin contribution, the transition to the trivial state makes the entropy jump to $\ln(6)\approx 1.8$, for $J=0$ (solid green), and to $\ln(3)\approx 1.1$, for finite $J$ (dashed green).
Here, the transition to $Z=0$ vanishes the hopping terms between atomic states, and the system is restricted to the manifold where the ground state lives, which is six-fold and three-fold degenerate for $J=0$ and $J>0$, respectively (see Appendix \ref{atomic-append}).
As expected, at $T=U/2=U_{c1}/4$ other atomic states start to be accessed by thermal fluctuations and the slave-spin entropy approaches $\ln(16)$ as $T$ is increased further.

In Fig. \ref{fig:entro-aniso} we plot the same quantities for the case when an OSBM occurs.
We have the same behaviour as for Fig. \ref{fig:entro-iso} respecting the FL phase ($T<T_\textrm{OSBM}$) and the trivial phase ($T>T_\textrm{coh}$).
For the intermediate OSBM phase, the narrow orbital collapse, $Z_2=0$, while the wide one remains metallic.
This explains the $\ln(4)$ jump in the fermionic entropy (blue), where the narrow orbital behaves as a free fermion gas ($Z_2=0$ implies infinite-temperature behaviour in orbital $2$).
The slave-spin contribution starts to grow in the OSBM phase, but it remains very small.

\section{Concluding remarks}
\label{sec:conclusion}

In conclusion, we have used the $Z_2$ slave-spin mean-field method to study the two-band Hubbard system at finite-temperature in the presence of Hund’s rule coupling and band anisotropy.
We have developed a finite-$T$ extension of the single-site approximation of the zero-$T$ formulation, that reproduces the physical limit for the uncorrelated case.
We have identified the temperature where the first-order transition between finite-$Z$ to $Z=0$ solutions occurs with the coherence temperature $T_\textrm{coh}$ that signals the crossover to a bad metal regime with incoherent quasiparticles.
When orbitals have different bandwidths, we have found a first-order transition to a phase where the quasiparticle weight of the narrow band vanishes ($Z=0$), the orbital-selective bad metal phase.
This intermediate phase between FL and bad metal phases is enhanced by the Hund's rule coupling, and its behaviour with a further increase in temperature can be related to a ferromagnetic Kondo-Hubbard lattice model.\cite{Biermann2005}
As expected, an increase in the Hund's rule coupling increase correlations, reducing the interorbital charge fluctuations, but increases the interorbital spin fluctuations. We highlight the qualitative difference between the $J=0$ and $J>0$ case, noting that it can be understood in term of the energy and degeneracy of the low-energy atomic configurations.

\section{Future directions}

From the point of view of the method, there are several improvements to the single-site SSMF that could be explored.
The freedom on the phase of the $c$-parameter allows exploration of the effects of using a complex quantity.
Also, a complex $c$-parameter becomes mandatory when performing a cluster mean-field approximation on the model.\cite{Hassan2010}
Other studies that utilise different slave-spin variants use the Schwinger boson representation to solve the quantum slave-spin Ising model,\cite{Ruegg2010,Yu2012} or construct a path-integral formulation that allows to perform Gaussian corrections to the single-site mean-field.\cite{Zhong2012b}
Recent calculations benchmark a variant of the $Z_2$ SSMF against the two-site Hubbard model, showing that slave-spin methods reproduce the exact behaviour of the ground state at half-filling, but also that special care has to be taken when moving away from the particle-hole symmetry, in which case the unphysical states have a big impact on the results.\cite{Yang2018}
Also, in a recent work a general formalism for slave-spin has been introduced\cite{Komijani2018} that reproduces the holon-doublon peak found in the two-band Hubbard model with accurate DMFT calculations.\cite{Nunez-Fernandez2018}

An interesting question to investigate in the future is to what extent the increase in the number of orbitals modifies the stability of the different phases of the Hund's metal at finite temperature.
It is suggested that the Hund's physics is more pronounced with increasing the number of orbitals.\cite{Fanfarillo2015}
For three orbitals or more, the Hund's rule acts for some commensurate fillings in an antagonistic ``Janus-faced'' manner, driving the system away from the Mott insulating phase while making the metallic phase more correlated.\cite{DeMedici2011a,Georges2013a}
DMFT with numerical renormalisation group calculations in the three-band model with two electrons ($1/3$ filling) show that spin-orbital separation is a generic feature of these systems, and that spin screening occurs at a much smaller energy scale than orbital screening or any other bare atomic excitation scale.\cite{Stadler2015,Stadler2018,Deng2019}
In a future study, we plan to extend the finite temperature SSMF method to more orbitals and away from half-filling (especially conmensurate fillings), investigating how $T_\textrm{coh}$ is modified by Hund's rule and the number of orbitals, the ``Janus-faced'' behaviour, spin-orbital separation, and the ``spin-freezing'' crossover.\cite{DeMedici2011a,Werner2008}

Our results show qualitative agreement with finite temperature DMFT calculations in the two-band model,\cite{Liebsch2004,Biermann2005,Knecht2005,Liebsch2005,Inaba2005,Liebsch2006,Inaba2007}
and future DMFT calculations should provide more precise test for our predictions, in particular the appearance of an OSBM phase and its dependence on $J$ and orbital anisotropy.
DMFT calculations with realistic band structures find that the coherence temperature is different for different bands in Sr$_2$RuO$_4$,\cite{Mravlje2011} and support the result of a temperature-induced coherent-incoherent crossover found experimentally in LiFeAs\cite{Miao2016a} and KFe$_2$Se$_2$.\cite{Yang2017a}
A systematic study of the coherent-incoherent crossover should be done regarding the different signatures of bad metallic behaviour, such as:
a very small crossover temperature (compared with other bare energy scales),
an increase of resistivity with temperature to rather large values,
an orbital-selective depletion of the spectral weight,
and the partial collapse of the Drude peak in the optical conductivity along with the transference of that spectral weight to higher frequencies.
Moreover, STM measurements of quasi-particle scattering interference should show a significant temperature dependence near the crossover to the orbital-selective bad metal.

\section{Acknowledgements}

We thank E. Bascones, L. Fanfarillo, G. Kotliar, F. Lechermann, L. de'  Medici, H. L. Nourse, L. Oberg and B. J. Powell for useful discusions.
This work was supported by an Australian Research Council Discovery Project, Grant No. DP160102425.

\appendix
\section{Finite temperature gauge parameter}
\label{opc-append}

For zero temperature, Hassan and de' Medici calculated in Ref. [\onlinecite{Hassan2010}] a choice for the parameter $c$ in the one band case. At the one-site mean-field level and restricting to real numbers, the expression $c=\frac{1}{\sqrt{\frac{n}{2}\, \left( 1 - \frac{n}{2}\right)}} -1$ reproduces the uncorrelated limit with $Z=1$, and only depends on the occupation number $n$ of the band.\cite{Hassan2010,DeMedici2017a}

We follow the steps and notation of Appendix A of Ref. [\onlinecite{Hassan2010}], and extend the calculations to finite temperatures, obtaining for the expectation values of $\hat{S}^z$ and $\hat{O}$,
\begin{align}
\langle \hat{O} \rangle =& - \frac{c\, a^* + a}{2\, R}\, \tanh \left( \beta\, R \right) \\
\langle \hat{S}^z \rangle =& - \frac{\lambda}{4\, R}\, \tanh \left( \beta\, R \right),
\end{align}
where $a=h+c\, h^*$ and $R=\sqrt{\frac{\lambda^2}{4} + \vert a \vert^2 }$. These quantities have to satisfy the self-consistent equations $Z= \langle \hat{O}^\dag \rangle\, \langle \hat{O} \rangle$ and $\langle \hat{S}^z \rangle=\frac{n}{2}-\frac{1}{2}$. Assuming a real $c$, we have $h=\sqrt{Z}\, \overline{\varepsilon}$ and $a=\sqrt{Z} \overline{\varepsilon}\left( 1+c\right)$, leading to the coupled equations,
\begin{equation}
\frac{ \tanh \left( \beta\, R \right)}{2\, R} = \frac{1}{-\overline{\varepsilon} \left(1+c\right)^2}=\frac{1+n}{\lambda}\ . \label{self-eqs-uncorr}
\end{equation}
In the uncorrelated limit $U=0$ we set the physical solution $Z=1$, and we determine the $c$ parameter by solving the equation,
\begin{align}
&\tanh \left( \beta\, \frac{ -\overline{\varepsilon}_0\, (1+c)^2}{2} \sqrt{ (1-n_0)^2 + \frac{4}{(1+c)^2}}\, \right)\nonumber \\
=&\ \sqrt{ (1-n_0)^2\, + \frac{4}{(1+c)^2}}\ . \label{opc}&
\end{align}
Also, the Lagrange multiplier is,
\begin{equation}
\lambda_0 = -\overline{\varepsilon}_0\, (1-n_0) (1+c)^2\ . \label{lambda0}
\end{equation}
Here, the subscript ``0'' refers to the calculation of quantities in the uncorrelated limit and at temperature $T$, i.e., using the occupation $n(\varepsilon)=\left( 1+e^{\beta (\varepsilon  + \epsilon_m - \mu)} \right)^{-1}$, where the omission of $\lambda$ in $n(\varepsilon)$ relates to the physical limit.
The SSMF yields a non-zero $\lambda$ in the uncorrelated limit,\cite{DeMedici2014, DeMedici2017, DeMedici2017a}
an unwanted behaviour that is solved by shifting $\lambda$ to satisfy the physical non-interacting limit $\lambda=0$.
Previous works use a numerical calculation of $\lambda_0$ to perform this shift.
Here, for each temperature, we solve first Eq. \eqref{opc}, then use the analytic formula \eqref{lambda0}, and insert the shifted quantity $\lambda - \lambda_0$ throughout the calculations.

The parameter $c$ and the shift $\lambda_0$ depend now on the occupation $n_0$, but also on the temperature $T$ and the non-interacting kinetic energy of the electrons $\overline{\varepsilon}_0$ (which exclusively depends on the filling and the shape of the bare density of states).
It is easy to check that Eq. \eqref{opc} recovers the known formula for $c$ at $T=0$, and also that $\lambda_0(T=0)=-4\, \overline{\varepsilon}_0 \frac{1-n}{n\, (2-n)}$. Also, at half-filling, we recover the physical value $\lambda_0=0$, for all temperature.

With the use of the parameter $c$ obtained from Eq. \eqref{opc}, $Z=1$ satisfy the one-band self-consistent equations at any temperature when $U=0$.
For the application to multiband systems, we use the same approach as for one band.
The non-interacting limit is just a set of uncoupled one-band systems, and each orbital $m$ has its $c_m$ and $\lambda_{m,0}$ determined by solving Eq. \eqref{opc} and \eqref{lambda0} for a particular $T$, occupation number $n_{m,0}$ and kinetic energy $\overline{\varepsilon}^{(m)}_0$.

\section{Construction of the physical solution}
\label{solutions-append}

\begin{figure}
  \includegraphics*[width=0.46\textwidth]{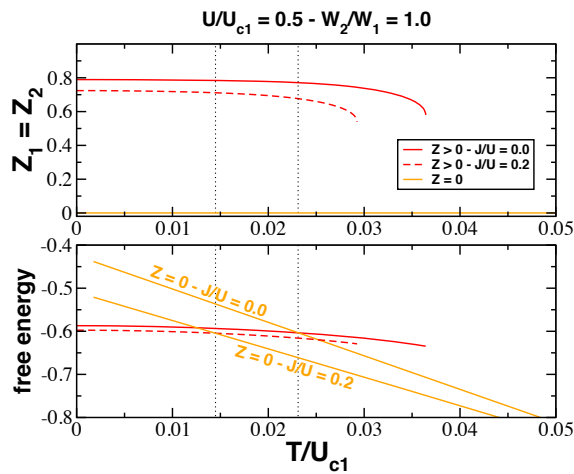}
  \caption{\label{fig:solutions-iso} Construction of the physical solution in the isotropic case.
  {\bf Top:} Several solutions to the self-consistent Eqs. \eqref{self-1} and \eqref{self-2} for the isotropic case $W_2=W_1$ and $J/U= 0$ and $0.2$.
  {\bf Bottom:} The corresponding free energy value of each solution (same colours used).
  The temperature where each finite-$Z$ solution free energy crosses the $Z=0$ one is the corresponding $T_\textrm{coh}$ (dotted lines).
  The solutions shown in the main text are constructed by concatenating the solutions with the lower free energy on each temperature range.}
\end{figure}

For the construction of the physical solutions, we explore the family of solutions to the self-consistent Eqs. \eqref{self-1} and \eqref{self-2}.
We have to remember that these solutions extremise the free energy (as a function of the mean-field parameters), but we still need to choose the solution that minimises it at each temperature.
As an example, in Fig. \ref{fig:solutions-iso}
 we show the construction of the physical solutions of Fig. \ref{fig:n100-Z-Tcoh} corresponding to $J/U = 0.0$ and $0.2$.
On the top, we show the family of solutions for each case, being the solution with finite $Z_1 = Z_2$ and the one with $Z_1=Z_2=0$. The dotted lines show the temperatures at which the free energy of the $Z=0$ solution becomes lower than the free energy for $Z>0$ (bottom), signalling a first-order transition from the later to the former.

\begin{figure}
  \includegraphics*[width=0.46\textwidth]{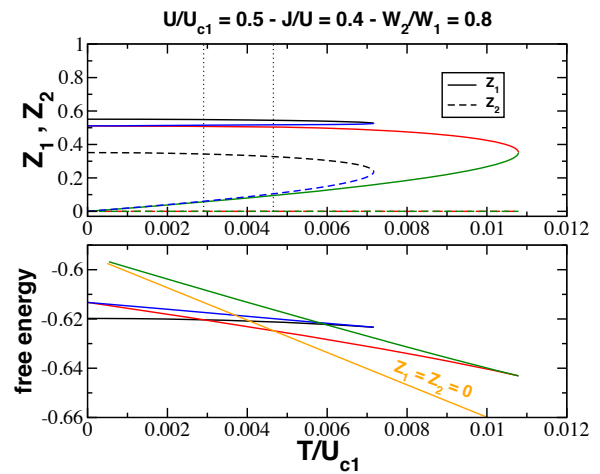}
  \caption{\label{fig:solutions-aniso-2} Construction of the physical solution in the anisotropic case.
  Each colour is a different solution.
  {\bf Top:} Several solutions to the self-consistent Eqs. \eqref{self-1} and \eqref{self-2} for $W_2=0.8\, W_1$, $U = 0.5\, U_\textrm{c1}$ and $J=0.4\, U$. Solid and dashed lines are $Z_1$ and $Z_2$, respectively.
  {\bf Bottom:} The corresponding free energy value of each solution (same colours used).
  Two transition temperatures exist in this case, namely $T_\textrm{OSBM}$ and $T_\textrm{coh}$.
  The former signals the first-order transition to the OSBM red solution, while the latter the transition to the $Z_1=Z_2=0$ trivial state (orange).
  The solutions shown in the main text are constructed similarly to the isotropic case, concatenating those with the lower free energy.}
\end{figure}

The same method is used when we have orbital anisotropy, obtaining this time a larger family of solutions of the self-consistent equations, as we can see on top of Fig. \ref{fig:solutions-aniso-2}.
\noindent Here, each colour means a different solution, and solid and dashed lines state for $Z_1$ and $Z_2$, respectively. Again, dotted lines mark the temperatures where a first-order transition occurs between black and red solutions, and red and orange (trivial) ones. The red solution, where $Z_1$ remains positive but $Z_2=0$, is what we call the orbital-selective bad metal (OSBM) phase.

\subsection{Limitations of the method}

\begin{figure}
  \includegraphics*[width=0.46\textwidth]{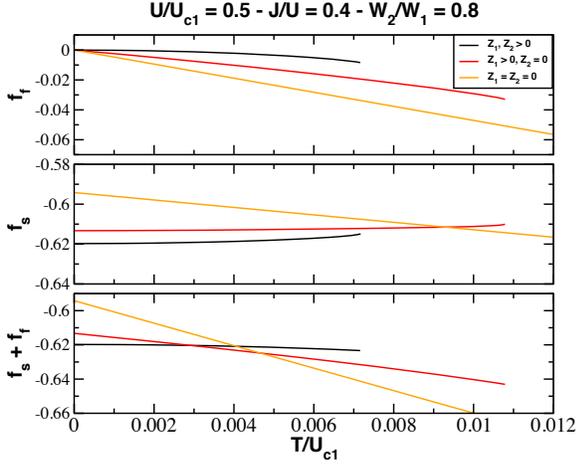}
  \caption{\label{fig:f-separated-aniso} Different contributions to the total free energy.
  For $W_2=0.8\, W_1$, $U = 0.5\, U_\textrm{c1}$ and $J=0.4\, U$, we plot the fermionic ({\bf top}), slave-spin ({\bf middle}), and total ({\bf bottom}) free energy (see main text).
  We use the same colours as in Fig. \ref{fig:solutions-aniso-2} for the three solution with the lowest free energy: renormalised Fermi liquid with $Z_1$ and $Z_2$ finite (black), OSBM with $Z_1$ finite and $Z_2=0$ (red), and bad metal with $Z_1=Z_2=0$ (orange).}
\end{figure}

We know that this method has limitations for the non-interacting limit $U=J=0$, which we discuss first in the interacting case of Fig. \ref{fig:solutions-aniso-2} for simplicity.
The total free energy of Eq. \eqref{free-energy} is the quantity minimised to find the final solution on each case, where the gradient is zero and $f$ has a minimum value.
In the general case, we can separate this free energy as $f=f_f + f_s$, where
\begin{align}
f_f =& \ - \frac{2}{\beta} \sum_m \int_{-\infty}^{\infty} \rho_m(\varepsilon)\, \ln \left( 1+e^{-\beta \left(Z_m \varepsilon -\mu -\lambda_m \right)}\right) \textrm{d}\varepsilon  - E_\textrm{MF} \label{f-fermionic} \\
f_s =& \ - \frac{1}{\beta} \ln \left( \mathcal{Z}_1^s \right) - E_\textrm{MF} \label{f-slavespin}\ .
\end{align}
Here, the fermionic part \eqref{f-fermionic} corresponds to the free energy of non-interacting fermions where each band has a bandwidth $Z\times W$. The quantity $E_\textrm{MF}$ shift the zero energy level to the Fermi surface.
The $Z_1=Z_2=0$ solution corresponds to a non-interacting flat band with $f=-T \ln (16)$, while any increase in the renormalisation $Z_m$ adds dispersion to the bands and increases the value of the free energy (top of Fig. \ref{fig:f-separated-aniso}).
Because of the slave-spin mapping, all the complexity of the original model goes exclusively into the slave-spin Hamiltonian, Eqs. (\ref{ss-hamiltonian-0}-\ref{ss-hamiltonian-J}).
In this sense, the transition between solutions are driven by the slave-spin contribution, as we can see in the middle of Fig. \ref{fig:f-separated-aniso}.
The effect of the fermionic contribution $f_f$ to the total free energy $f$ is to reduce the transition temperatures observed in the slave-spin contribution $f_s$ (bottom of Fig. \ref{fig:f-separated-aniso}).

\begin{figure}
  \includegraphics*[width=0.46\textwidth]{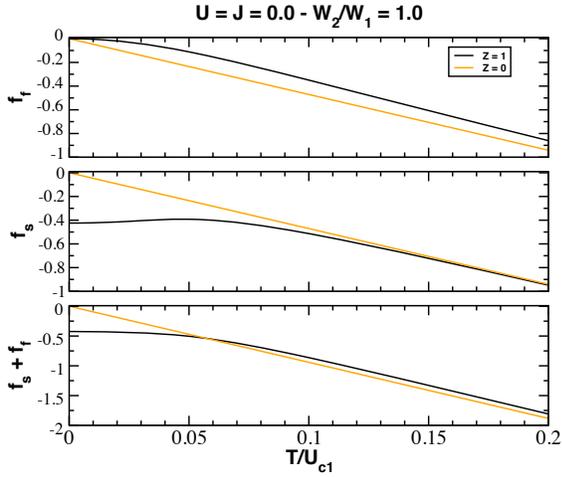}
  \caption{\label{fig:f-separated-iso} Different contributions to the total free energy in the non-interacting, isotropic case. ({\bf top}) fermionic contribution, ({\bf middle}) slave-spin contribution, and ({\bf bottom}) total free energy.
  The slave-spin contribution to the free energy for the $Z=1$ solution is lower than the corresponding to $Z=0$ for all temperature. The addition of the fermionic contribution $f_f$ to obtain the total free energy results in a crossing of the free energy of both solutions.}
\end{figure}

Regarding the non-interacting limit $U=J=0$, shown in Fig. \ref{fig:f-separated-iso},
even though the slave-spin free energy contribution for the $Z=1$ solution is lower than the $f_s$ for $Z=0$ at all $T$, the opposite effect on the fermionic contribution $f_f$ results in an unphysical crossing between the total free energy of the $Z=1$ and $Z=0$ solutions at a finite temperature.

\section{Atomic states of the two-band system}
\label{atomic-append}

\begin{figure}
  \includegraphics[width=0.46\textwidth]{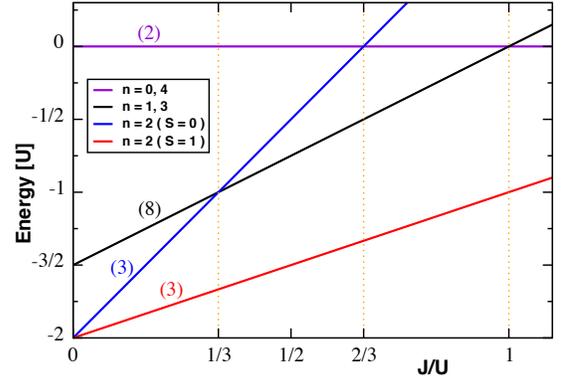}
  \caption{\label{fig:energies-hf} Dependence of the energies of the atomic configurations in function of $J/U$. The numbers in parentheses denotes the degeneracy of the state.
  For $J=0$ we can see that the $S=0$ and $S=1$ sectors are degenerated.
  This degeneracy is lifted as soon as $J$ becomes finite, and the $S=1$ sector becomes the lowest in energy.
  The next in energy states are those corresponding to the $S=0$ sector for $0<J<U/3$, and those corresponding to the total spin $S=\frac{1}{2}$ for $U/3<J<U$.}
\end{figure}

We can achieve a good understanding of the physics underlying the system by looking at the possible atomic states and its energies, i.e., the eigenstates of the local Hamiltonian terms, $\hat{\mathcal{H}}_{J} +\hat{\mathcal{H}}_{U} -\mu\, (\hat{n}_1+\hat{n}_2)$.
Using Eqs. \eqref{HU} and \eqref{HJ}, and the known value of the chemical potential at half-filling $\mu = E_0 = \frac{3\, U}{2} - \frac{3\, J}{2}$, we show in Table \ref{atomic-states} the energy and total spin of these states.
\begingroup
\squeezetable
\begin{table}
\caption{\label{atomic-states} Atomic states in absence of hopping hybridization, with its corresponding energy, electron number and total spin per site.
Different colour clusters same energy states, and correspond with the colours used in Fig. \ref{fig:energies-hf}. Last column list the energy at half-filling (h-f) for the atomic states.}
\begin{tabular}{|c|c|c|c|c|c|}
\hline
State & $n_1\! +\! n_2$ & Total spin & Energy & Energy (h-f) \\ \hline \hline
\cellcolor{violet!50} $\vert \ O_1 ; \ O_2 \rangle$ & $0$ & $0$ & $0$ & $0$ \\ \hline
\cellcolor{black!40} $\vert \Uparrow_1 ; \ O_2 \rangle$ & & & & \\ \cline{1-1}
\cellcolor{black!40} $\vert \Downarrow_1 ; \ O_2 \rangle$ & $1$ & $\frac{1}{2}$ & $0 -\mu$ &  $-\frac{3\, U}{2} + \frac{3\, J}{2}$ \\ \cline{1-1}
\cellcolor{black!40} $\vert \ O_1 ; \Uparrow_2 \rangle$ & & & & \\ \cline{1-1}
\cellcolor{black!40} $\vert \ O_1 ; \Downarrow_2 \rangle$ & & & & \\ \hline
\cellcolor{blue!50} $\vert \Uparrow \Downarrow_1 ; \ O_2 \rangle$ & & & & \\ \cline{1-1}
\cellcolor{blue!50} $\vert \ O_1 ;  \Uparrow \Downarrow_2 \rangle$ & $2$ & $0$ & $ U -2\, \mu$ & $-2\, U + 3\, J$ \\ \cline{1-1}
\cellcolor{blue!50} $\frac{\vert \Uparrow_1 ; \Downarrow_2 \rangle - \vert \Downarrow_1 ;\Uparrow_2 \rangle}{\sqrt{2}}$ & & & & \\ \hline
\cellcolor{red!50} $\vert \Uparrow_1 ; \Uparrow_2 \rangle$ & & & & \\ \cline{1-1}
\cellcolor{red!50} $\vert \Downarrow_1 ; \Downarrow_2 \rangle$ & $2$ & $1$ &  $U - 2\, J -2\, \mu$ & $-2\, U + J$ \\ \cline{1-1}
\cellcolor{red!50} $\frac{\vert \Uparrow_1 ; \Downarrow_2 \rangle + \vert \Downarrow_1 ;\Uparrow_2 \rangle}{\sqrt{2}}$ & & & & \\ \hline
\cellcolor{black!40} $\vert \Uparrow \Downarrow_1 ; \Uparrow_2 \rangle$ & & & & \\ \cline{1-1}
\cellcolor{black!40} $\vert \Uparrow \Downarrow_1 ; \Downarrow_2 \rangle$ & $3$ & $\frac{1}{2}$ & $3\, U -3\, J -3\, \mu$ & $-\frac{3\, U}{2} + \frac{3\, J}{2}$ \\ \cline{1-1}
\cellcolor{black!40} $\vert \Uparrow_1 ;  \Uparrow \Downarrow_2 \rangle$ & & & & \\ \cline{1-1}
\cellcolor{black!40} $\vert \Downarrow_1 ;  \Uparrow \Downarrow_2 \rangle$ & & & & \\ \hline
\cellcolor{violet!50} $\vert \Uparrow \Downarrow_1 ;  \Uparrow \Downarrow_2 \rangle$ & $4$ & $0$ & $6\, U - 6\, J -4\, \mu$ & $0$ \\ \hline
\end{tabular}
\end{table}
\endgroup
We colour the states in four groups regarding their atomic energies and plot its evolution with $J/U$ in Fig. \ref{fig:energies-hf}.
Here the energy is in units of $U$, and we also write in parentheses the degeneracy of each energy state.
For $J=0$ the lowest (atomic) energy states live within the six-fold manifold involving the $S=0$ and $S=1$ sectors, while a finite $J$ lifts this degeneracy into the two spin sectors, making the $S=1$ triplet the lowest energy sector.\cite{Komijani2017}
This change in the degeneracy of the lowest energy sector whether $J=0$ or $J>0$ change qualitatively the behaviour of the ground state,\cite{Koga2004,DeMedici2017a} and is expected to also affect the low temperature properties.
Another relevant energy scale can be $J=\frac{U}{3}$, where the energy of the $S=0$ and $S=\frac{1}{2}$ sectors crosses, signalling a possible qualitative change in the low energy excitations.


\end{document}